\documentclass{aa}  
\usepackage{natbib}
\usepackage[varg]{txfonts}
\usepackage{tikz}
\usepackage{ulem}
\usetikzlibrary{shapes,arrows, decorations.pathmorphing, decorations.markings}
\usepackage{graphicx}
\usepackage{multirow}

\newcommand{\Cheops}{{CHEOPS}}
\newcommand{\DRP}{{DRP}}

\begin{document} 
	
	\title{Expected performances of the Characterising Exoplanet Satellite (CHEOPS)}
	\subtitle{III. Data reduction pipeline: architecture and simulated performances.}
	%\title{\Cheops\ data reduction pipeline: architecture and simulated performances}                                                          
	
	\author{S.Hoyer
		\inst{1}
		\and
		P.Guterman\inst{1,2}
		\and
		O.Demangeon\inst{1,3}
		\and
		S.G.Sousa\inst{3}
		\and
		M.Deleuil\inst{1}
		\and
		JC.Meunier\inst{1}
		\and
		W. Benz\inst{4}
	}
	
	\institute{Aix Marseille Univ, CNRS, LAM, Laboratoire d'Astrophysique de Marseille, Marseille, France
		\and Division Technique INSU, BP 330, 83507 La Seyne cedex, France
		\and Instituto de Astrof\'isica e Ciencias do Espa\c co, Universidade do Porto, CAUP, Rua das Estrelas, Porto 4150–762, Portugal
		\and University of Bern, Physikalisches Institut, Gesellschaftstrasse 6, CH-3012, Bern, Switzerland
	}

	\date{Received: July 16, 2019; accepted: September 16, 2019 }
	
	% \abstract{}{}{}{}{} 
	% 5 {} token are mandatory
	
	\abstract{ The CHaracterizing ExOPlanet Satellite (\Cheops), to be launched in December 2019, will detect and characterize small size exoplanets via ultra high precision photometry during transits. \Cheops\ is designed as a follow-up telescope and therefore it will monitor a single target at a time. The scientific users will retrieve science-ready light curves of the target, automatically generated by the \Cheops\ data reduction pipeline of the Science Operations Centre. This paper describes how the pipeline processes the series of raw images and, in particular, how it handles the specificities of \Cheops\ data, such as the rotating field of view, the extended irregular Point Spread Function, and the data temporal gaps in the context of the strict photometric requirements of the mission. The current status and performance of the main processing stages of the pipeline, that is the \textit{calibration}, \textit{correction} and \textit{photometry}, are presented to allow the users to understand how the science-ready data have been derived. Finally, the general performance of the pipeline is illustrated via the processing of representative scientific cases generated by the mission simulator. }

	
	\maketitle
	%
	%________________________________________________________________
	
	\section{Introduction}

\Cheops\ (CHaracterizing ExOPlanet Satellite) is an ESA small mission to be launched in December of 2019. \Cheops\ is designed as a follow-up instrument devoted to ultra high precision photometry, able to detect and/or precisely measure transits of small size exoplanets already known via radial velocity measurements or via transit searches \citep{Broeg2014, Benz2018}. The series of raw images acquired by the instrument will be automatically processed (with no external interaction nor interactive configuration) into a flux time series, ready for scientific analyses. As part of the Science Operations Centre, the Data Reduction Pipeline (\DRP) is in charge of producing these calibrated light curves, with associated intermediate products, which will be delivered to the scientific users. While the instrument performs ultra-high precision photometry like the CoRoT \citep{Deru2015, Corot2016} or Kepler \citep{Jenkins2010a, Jenkins2010b} missions, it presents different specificities that demand tailored approaches for the data reduction. In particular, the instrument field of view is rotating around the line of sight, making background stars to roll around the target, and potentially periodically polluting its photometry. In addition, the Point Spread Function (PSF) of the instrument measured in laboratory exhibits an extended irregular shape which, together with the temporal gaps in the data, challenges pipeline procedures such as the detection and correction of cosmic rays hits, among others.

The present paper aims at providing to the community a complete description of the automated data reduction pipeline, as implemented in the
pre-launch phase. The goal is to show how the pipeline deals with \Cheops\ specificities, allowing a better understanding of how the science-ready data have been derived. Also, this paper intends to serve as reference for the possible use of additional pipeline products which will complement further light curve analysis (e.g. with filtering or detrending algorithms). Finally, in the framework of the specific and strict mission photometric requirements, the expected performance of the \DRP\ have been estimated. For this, series of simulated data for typical astrophysical configurations, as provided by \Cheops\ end-to-end simulator (Futyan et al. \textit{subm.}), have been used. The reader can also find a description of the \Cheops\ on-ground performance in Deline et al. (\textit{accepted}).

The structure of the paper is the following: Sect.~\ref{sec:mission} recalls the mission profile and instrument specificities while Sect.~\ref{sec:archi} presents the pipeline architecture. Sections~\ref{sec:cal} to \ref{ssec:phot} details the different processing steps operated by the main modules of the \DRP\ and the implemented algorithms. Some processes which could be used indistinctly at any step of the pipeline are described in Sect.~\ref{sec:Gen}. The expected performance are reviewed in Sect.~\ref{sec:performance} and Sect.~\ref{sec:conclusions} summarizes and concludes this work.

\section{Overview of the mission profile and instrument} 
    \label{sec:mission}
	
A complete description of the instrument and the mission profile can be found in the \Cheops\ \textit{Observers Manual\footnote{https://www.cosmos.esa.int/web/cheops-guest-observers-programme/ao-1}} or in \cite{Broeg2018} and  Deline et al. (\textit{in press}).
But to make reading easier, we provide a quick description of the key elements which direct the data reduction in the following paragraphs.  
\Cheops\ will be settled on a $700$ km altitude sun-synchronous orbit of about $101$ minutes. The spacecraft is nadir locked and will continuously roll around the line of sight, ensuring a thermally stable environment for the payload radiators. As a consequence, during one orbit the background stars rotate around the optical axis of the telescope while the target star remains at the same location, modulo jitter perturbations. Because of its orbit at low altitude, it is expected that up to $40\%$ of data could be lost due to the passage of the Earth too close to the line of sight for targets far from the ecliptic but also due to the SAA (South Atlantic Anomaly) crossings \citep{PdaSilva2008}. Therefore, these losses translate into time gaps in the raw data products received by the \DRP\ and consequently, in the final light curves delivered by the pipeline.
As mentioned before, the telescope will observe one single target at a time in a field of view of 0.32$\degr$ in diameter. The telescope has an effective diameter of 30 cm, it has no shutter and the focal plane is equipped with a 1024$\times$1024 pixels back-illuminated CCD with a pixel size of 13~$\mu$m and a pixel scale of 1\arcsec. It will operate at a nominal temperature of -40\degr~C. The focal plane is defocussed to deliver a large Point Spread Function (PSF) with a 12 pixels radius encircling 90\% of the flux. As a result of the combination of the Ritchey-Chr\'etien design and other specific features of the building of the telescope, the PSF exhibits sharp and peaky features at sub-pixel level. \Cheops\ has no filter in the optical path and its bandpass covers the visible-near-infrared range of 330-1100 nm. The spectral transmission of CHEOPS is very similar to the Gaia G bandpass \citep{gaiadr2}, see Fig. 1 in Deline et al. (\textit{in press)}. At launch, the telescope will have a cover for protection. The opening of the cover will occur during in-orbit commissioning after some tests and calibration observations of the instrument.  

Full-array images will be downloaded for calibration or test purposes only. In nominal operation mode only an image of 200$\times$200 pixels (default size), referred as subarray, will be downlinked to ground with the associated housekeeping data. Each subarray image, usually is formed by the stack of several shorter exposures which allows to, for example, avoid saturation during bright target observations. Complementary, the central region of the image is transmitted before the stacking as small imagettes of typical size 35$\times$35 pixels, providing thus a higher cadence sampling of the target. In fact, images and imagettes are circular in order to downlink only the relevant region of the images and thus spare bandwidth.

The magnitude of targets are in the 6$\leq$V-mag$\leq$12 range but the instrument will also allow the observation of brighter or much fainter stars. To accommodate this large range of magnitudes the exposure time can be adjusted from 1~ms to 60~s. Depending on the selected exposure time, different detector read-out modes are set-up. These modes, called \textit{faint}, \textit{faint-fast}, \textit{bright} and \textit{ultra-bright}, consist of different read-out frequencies and different setting combinations of the detector read-out and the on-board processing of the image. Thus, each read-out mode has an specific duty cycle (as low as 8-50$\%$ for exposure times below 1~s) which leads finally to an image cadence between 1~s and 60~s (see Table 1 of \Cheops ~\textit{Observers Manual} for details). In addition, in order to reduce the amount of downlinked data, one image can be the stack of 1 to 60 short exposures. 

The instrument is required to reach a photometric precision of 20 ppm for a star with a V magnitude in the range  $6\leq V$-mag$ \leq9$ with 6 hours of integration time, to allow the detection of an Earth-like planet around a G5V star with an orbital period of 50 days. At the faint end, the expected photometric precision is 85~ppm for a star of $V$-mag=12 in 3 hours of integration time, which will allow the detection of Neptune-size planets transiting a K-type dwarf star  with an orbital period of 13 days \citep{Fortier2015, Benz2018}. To achieve the high photometric stability, the instrument should operate in a thermally stable environment, minimize the various sources of straylight, and ensure a pointing stability of 2\arcsec~rms. This precision will be achieved by including the instrument in the attitude control loop.

\section{Pipeline architecture} 
\label{sec:archi}

The \DRP\ is run automatically once triggered by the processing framework. There is no interaction with external agents and there is no interactive configuration of the pipeline. 

The complete processing can be separated in 3 main steps: 1) the \textit{calibration} module which corrects the instrumental response, 2) the \textit{correction} module on charge of correcting environmental effects and 3) the \textit{photometry} module which transforms the resulting calibrated and corrected images into a calibrated flux time series or light curve. Each of these modules consist in successive processing steps (Fig.~\ref{fig:Flowchart}) which are run sequentially as the output of one step is used as an input of the next one. The next sections detail the  different processing steps and the adopted algorithms of each of these 3 main modules. Some additional modules which are used indistinctly at any point of the pipeline are described in Sect.~\ref{sec:Gen}. 

In addition to the reduced light curves, the pipeline generates a visit processing report. This report allows the user to get direct insight into the performance of each step of the data reduction. 
	
	% Define block styles
	\tikzstyle{roundblock} = [rectangle, draw=gray, fill=gray!10, 
	text width=6em, text centered, rounded corners=1em, minimum height=1em]
	\tikzstyle{grectblock} = [rectangle, draw=green, fill=green!20, text width=7em, text centered]
	\tikzstyle{orectblock} = [rectangle, draw=orange, fill=orange!20, text width=7em, text centered]
	\tikzstyle{brectblock} = [rectangle, draw=blue, fill=blue!20, text width=7em, text centered]
	\tikzstyle{arrow} = [draw, -latex']
	\tikzstyle{line} = [draw]
	
	\begin{figure}
	\centering
		\begin{tikzpicture}[auto]
		% Calib
		\node [roundblock] (raw) {raw images};
		\node [grectblock, below of=raw] (eventflagging) {event flagging};
		\path [arrow] (raw) -- (eventflagging);
		\node [grectblock, below of=eventflagging] (bias) {bias};
		\path [arrow] (eventflagging) -- (bias);
		\node [grectblock, below of=bias] (gain) {gain};
		\path [arrow] (bias) -- (gain);
		\node [grectblock, below of=gain] (linearize) {linearize};
		\path [arrow] (gain) -- (linearize);
		\node [grectblock, below of=linearize] (darkcurrent) {dark current};
		\path [arrow] (linearize) -- (darkcurrent);
		\node [grectblock, below of=darkcurrent] (flatfield) {flat field};
		\path [arrow] (darkcurrent) -- (flatfield);
		
		% Correction
		\node [orectblock, right of=raw, xshift=7em] (smearing) {smearing};
		\node[inner sep=0,minimum size=0,right of=flatfield,xshift=2em] (k) {}; % invisible node
		\path [line] (flatfield) -- (k);
		\path [arrow] (k) |- (smearing);
		\node [orectblock, below of=smearing] (badpixel) {bad pixels};
		\path [arrow] (smearing) -- (badpixel);
		\node [orectblock, below of=badpixel] (depointing) {depointing};
		\path [arrow] (badpixel) -- (depointing);
		\node [orectblock, below of=depointing] (pixelphys) {pixel $\leftrightarrow$ sky};
		\path [arrow] (depointing) -- (pixelphys);
		\node [orectblock, below of=pixelphys] (background) {background \& straylight};
		\path [arrow] (pixelphys) -- (background);
		
		% photometry
		\node [brectblock, below of=background] (imagesimulation) {image simulator};
		\path [arrow] (background) -- (imagesimulation);
		\node [brectblock, below of=imagesimulation] (aperture) {aperture};
		\path [arrow] (imagesimulation) -- (aperture);
		\node [brectblock, below of=aperture] (photometry) {photometry};
		\path [arrow] (aperture) -- (photometry);
		\node [brectblock, below of=photometry] (contamination) {contamination};
		\path [arrow] (photometry) -- (contamination);
		\node [roundblock, below of=contamination] (lightcurve) {light curve};
		\path [arrow] (contamination) -- (lightcurve);
		\end{tikzpicture}
		\caption{Data reduction flowchart. Green, orange and blue color are \textit{calibration}, \textit{correction} and \textit{photometry} main modules, respectively.}
		\label{fig:Flowchart}
	\end{figure}
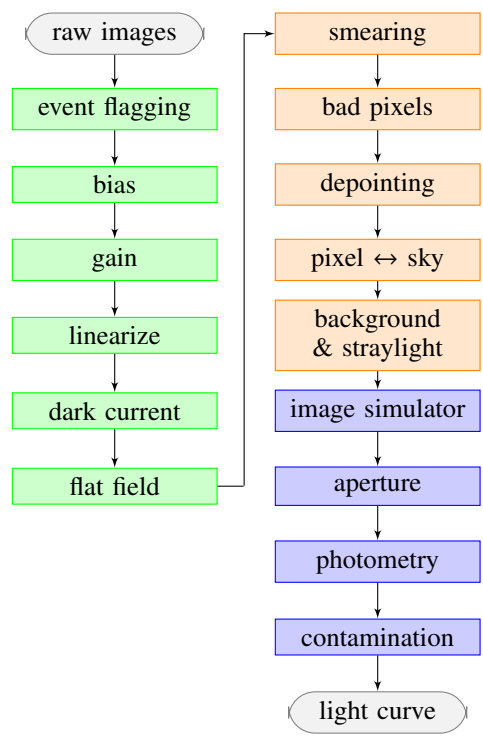

\section{Calibration}
    \label{sec:cal}
	% --------------------------
	
	The calibration step transforms the raw images received from the instrument, into photo-electrons calibrated ones. It exploits the knowledge of the instrument derived from its characterization performed either in laboratory or in space during the commissioning phase, to invert the instrument response. Thus, the calibration module removes the bias introduced by the analogue chain, restores the ADU (Analogue Digital Unit) image back to e$^{-}$, evaluates and corrects the dark current and the Pixel Response Non Uniformity (PRNU or flat-field).
	
	\subsection{Instrument model}
    \label{ssec:model} 
	The data reduction sequence results from the signal transforming steps from incident photons to raw images as illustrated in Fig.~\ref{fig:signalchain}.

	% Define flowchart block styles
	\tikzstyle{roundblock} = [rectangle, draw=gray, fill=gray!10, 
	text width=6em, text centered, rounded corners=1em, minimum height=1em]
	\tikzstyle{grectblock} = [rectangle, draw=green, fill=green!20, text width=7em, text centered]
	\tikzstyle{orectblock} = [rectangle, draw=orange, fill=orange!20, text width=7em, text centered]
	\tikzstyle{brectblock} = [rectangle, draw=blue, fill=blue!20, text width=7em, text centered]
	\tikzstyle{arrow} = [draw, -latex']
	\tikzstyle{line} = [draw]

	% --- image formation
	\begin{figure}
	\centering
		\begin{tikzpicture}[>=stealth',pos=.8,
		ylow/.style={yshift=-2cm},
		photon/.style={decorate,decoration={snake,post length=1mm}}]
		
		% Star --> focal plane
		\node [left] (target) {$f^\star_\text{ph}$};     % Star flux
		\node [right of= target] (tmp10) {};    % invisible center of arrow
		\node [above of= tmp10] (tmp11) {$T$};    % invisible top center of arrow
		\draw[->] (tmp11) -- (tmp10);
		\node [right of=tmp10] (focalplane) {[ph]}; % focal plane
		\node [below of=focalplane, yshift = 1.5em] {focal plane}; % label focal plane
		\draw[->,photon] (target) -- (focalplane);
		
		% Focal plane -> pixel
		\node [right of= focalplane] (tmp20) {};    % invisible center of arrow
		\node [above of= tmp20,align=center] (tmp21) {$Q~\times~F$ \\ $[ \text{e}^{-} \text{ph}^{-1}]$};  % invisible top center of arrow

		\node [right of=tmp20] (pixel) {[$\text{e}^{-}$]};
        \node [above of= pixel, yshift = -1.5em] {$f^\star_\text{e}$}; % text f_e^star above 'ccd'
		\node [below of=pixel, yshift = 1.5em] {ccd}; % label focal plane
		\draw[->] (focalplane) -- (pixel);
		\draw[->] (tmp21) -- (tmp20);
		
		% + dark
		\node[draw,circle,inner sep=0pt, right of=pixel] (addark) {+}; 	% "+" sign
		\node [right of=addark] (dark) {[$\text{e}^{-}$]};							% end point
		\node [above of= addark,text width=3em,,align=center] (tmp31) {$d_\text{e}$ \\ $[\text{e}^{-} s^{-1}]$};    % invisible above addition
		\node [below of= addark,text width=3em, yshift = 1.5em, align=center] (darklegend) {dark};    % invisible above addition
		\draw[-] (pixel) -- (addark);
		\draw[->] (addark) -- (dark);
		\draw[->] (tmp31) -- (addark);
		
		% readout
		\node [right of=dark,inner sep=0,minimum size=0,xshift=5mm] (tmp40) {};		 % U-turn horz
		\node [below of=tmp40, rotate=270, draw] (readout) {readout};  % U-turn half vert
		
		% gain
		\node [below of=dark, ylow,draw,regular polygon, regular polygon sides=3,shape border rotate=90,inner sep=.5mm,xshift=.7cm] (gain) {$g$};	
		\node [above of=gain, yshift = -1em, align=center] {$[V_{\text{L}} ~\text{e}^{-1}]$}; % label ampli
		\node [below of=gain, yshift = 1em, align=center] {gain}; % label ampli
		\draw (dark) -| (tmp40) -| (readout) |- (gain);
		
		\node [left of=gain,xshift=-1em] (volt) {[V$_{\text{L}}$]};
		
		% Non linearity
		\node [left of=volt,xshift=-.4em,rectangle,draw] (nlbox){NL};
		\node [below of=nlbox, yshift = 1.em, align=center] (nllegend) {non linearity};
		\node [left of=nlbox, xshift=-.4em] (nlout) {[V$_{\text{NL}}$]};
		\draw[->] (gain) -- (volt);
		\draw[->] (volt) -- (nlbox);
		\draw[->] (nlbox) -- (nlout);
		
		% Bias
		\node[left of=nlout, draw, circle, inner sep=0pt] (addbias) {+}; 	% "+" sign
		\node [above of=addbias, yshift = .5em, align=center] (bias) {$b_{\text{V}}$}; % bias
		\node [below of=addbias, yshift = 1.em, align=center] (biaslegend) {bias}; % label bias
		\draw[->] (bias) -- (addbias);
        \draw[->] (nlout) -- (addbias);

		% ADC
		\node [left of=addbias,draw, xshift=.6em] (adcbox){AD};        
        \node [above of=adcbox, yshift=-1.5em, align=center] (adctext) {[adu V$^{-1}_{\text{NL}}$]};
        
        % Image
        \node at (target|-adcbox) (Iadu) {$f^\star_\text{adu}$};
        \draw[->] (addbias) -- (adcbox) -- (Iadu);
		
		\end{tikzpicture}
		\caption{Signal chain. Following the main arrow, $f^\star_\text{ph}$ is the input photon flux. The units of successive transformations are given in brackets: [ph] photons, [e$^{-}$] electrons, $[\text{V}_\text{L}]$ and $[\text{V}_\text{NL}]$ linear and non linear volts, and $[\text{adu}]$ the analogue-to-digital units. $T$ is the optical throughput, $Q$ is the quantum efficiency, $F$ the flat field, $d_{\text{e}}$ is the dark current. The \textit{readout} label is the frame transfer, the triangle represents the analogue amplifier with its gain $g$, its non linearity \textit{NL} and its bias voltage $b_V$. AD is the analogue to digital converter. The output is the raw image $f^\star_\text{adu}$.}
		\label{fig:signalchain}
	\end{figure}
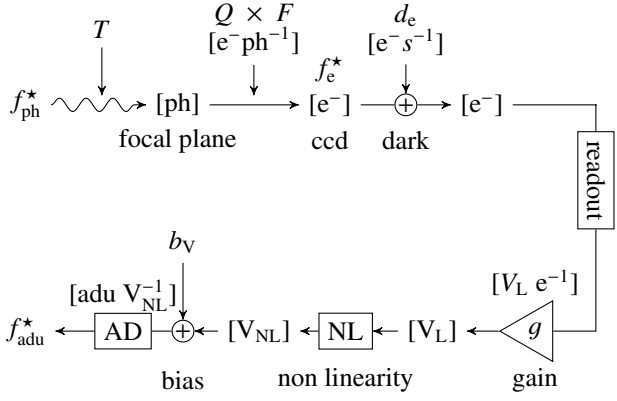

The light flux $f^\star_\text{ph}$ entering the telescope is guided to the focal plane trough the optics with an optical throughput $T$ that depends on the wavelength and incidence angle. The photons create opto-electrons in the CCD with a quantum efficiency rate $Q$ depending on the wavelength. 
Different response from pixel to pixel creates a pattern that translates into PRNU, which is evaluated as a function of wavelength in laboratory.  At the end of the exposure time, the frame is  transferred in 25 ms to the CCD storage zone protected from light. A dark current leakage of $\sim$0.05 e$^{-}$ pix$^{-1}$ s$^{-1}$ adds to the photo-electrons during both exposure and readout process. The pixels are then serialized and their charge converted into voltage by the analogue amplifier with gain $g$, deviation from linearity NL, and an added polarization bias voltage $b_V$ to prevent feeding the digital converter with possibly slightly negative voltage. The serialization lasts from $\sim$1~s to 4.63~s depending on the chosen reading mode. The result is the raw image in ADU received from the instrument. 
	
The overall transformation from star's photons to ADU is:
	\begin{equation}
		f^\star_\text{adu} = \text{AD}\Big(\text{NL} \Big[g \cdot(f^\star_\text{ph}\cdot T\cdot Q \cdot F+d_\text{e})\Big]+b_\text{V}\Big),
		\label{eq:f_adu}
	\end{equation}following the labelling of the different transformations presented in Fig.~\ref{fig:signalchain}. The flux received on the focal plane can be retrieved by inverting  Eq.\ref{eq:f_adu}:

\begin{equation}
	f^\star_\text{e} = \Big[
   \text{L}_\text{e}  \Big(  \frac{f^\star_\text{adu}-b_\text{adu}}{g} \Big)   - d_\text{e} \Big] \cdot \frac{1}{F},
\label{eq:f_ph}
\end{equation}where $f^\star_\text{e} = f^\star_\text{ph} \cdot T \cdot Q_{\text{e}~\text{ph}^{-1}}$ is the input flux in units of electrons, the function $\text{L}_\text{e} = [\text{AD}(\text{NL})]^{-1}$ is the inverse of the volts non linearity after digitization. This function is derived from laboratory measurements (Sect.~\ref{ssec:linearization}). The digitized bias voltage is $b_\text{adu}$ measured as explained in Sect.~\ref{ssec:bias}. The measurement of dark electrons $d_\text{e}$ is described in Sect.~\ref{ssec:dark}.
	
The organization of the data reduction pipeline presented in Fig.~\ref{fig:Flowchart} is derived directly from the signal restoration in Eq.~\ref{eq:f_ph}. The first step in the calibration module is the \textit{event flagging}, which is a general function of the pipeline responsible to flag images previous to any processing of data. This function is described in details in Sect.~\ref{ssec:event}.
	
\subsection{Bias and readout noise}
\label{ssec:bias}
	% --------------------------
	
The bias voltage is a voltage added to avoid negative values due to readout noise in case of faint flux. For \Cheops's CCD the bias voltage is regulated around $b_\text{adu}\sim$609 ADU per readout with a 10 ppm stability. The expected readout noise ($\text{ron}$) is $\sim$3.5 ADU per readout. Since the reference voltage used to generate the bias voltage can vary slightly with temperature, the bias is monitored and corrected using prescan pixels.
    
\begin{figure}
	\centering
		\includegraphics[width=\linewidth]{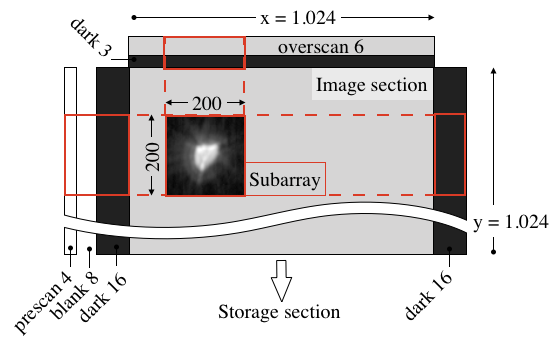}
		\caption{Schematic view of the photo-sensitive area of the CCD: the $200 \times 200$ square inside the 1024x1024 full CCD is the region of interest, called \textit{subarray}, transmitted to ground. Margins on the left: 4 prescan columns, 8 blank columns (unused), and 16 + 16 dark columns. Margins on the top: 6 overscan rows, 3 dark rows. The bottom storage section, not represented, mirrors the CCD, including margins. The arrows on the top/right of the diagram represent the x/y-axis of the pixels of the full CCD.}
		\label{fig:pixmap}
	\end{figure}

Prescan are virtual, empty pixels that contain neither photon nor dark current electrons and they are digitized before any real pixel. For prescan pixels, Eq.~\ref{eq:f_adu} simplifies to $b_\text{adu}=\text{AD}(b_\text{V})$. The CCD pixels map is presented in Fig.~\ref{fig:pixmap}, where the columns and rows correspond to the x- and y-axis, respectively. Prescans, take the form of 4 extra columns on this map.

To save bandwidth, only the median $\bar{p}$ and standard deviation $\sigma_p$ of the on board stacked prescan pixels are transmitted to ground. The pipeline then normalizes to a single readout value to estimate the bias $\hat{b}_\text{adu} = \bar{p}/n$ and readout noise $\hat{\text{r}}_o = \sigma_p/\sqrt{n}$ of a single image, where $n$ is the number of stacked exposures. 

In practice, using such bias estimate would cause a significant increase of the white noise in the light curve of the order of $(n_{\text{ap}}/\sqrt{n_{\text{presc}}}) \times \mathrm{ron}\sim100 \times \mathrm{ron}$, with $n_{\text{ap}}$ being of the order of 3000 pixels in the aperture (assuming a 30 pixels radius) and $n_{\text{presc}}$ the 800 prescan pixels.

To overcome this, the bias correction is separated into two steps: 1) a constant component is accurately estimated over the whole visit which can therefore be subtracted from images without creating noise. This constant component, <$\hat{b}_{\text{adu}}$>, is the average of $\hat{b}_\text{adu}$ over the visit; and 2) the time varying component is then corrected later by the general background correction (Sect.~\ref{ssec:bkg}) that works on an image per image basis. This component is assumed to be small thanks to the high thermal stability of the instrument.

Additionally, the bias difference between pixels is compensated by using a fixed bias frame recorded using null time exposures during ground calibration and updated regularly in flight. The pixel couplings that would result as image structures have actually been found to be negligible. The overall bias correction then resumes to: 
	\begin{equation}\label{eq_biasall} 
	I_b =  I_r - n\cdot\text{<}\hat{b}_\text{adu}\text{>} - n\cdot B_b,
	\end{equation}
with $I_r$ the raw image, $I_b$ the bias corrected image, $B_b$ the zero-average bias frame and $n$ the number of stacked images.

	\subsection{Gain}
        \label{ssec:gain}
	% --------------------------
	
The analogue amplifier converts the individual charges into a low impedance voltage that feeds the AD converter. The amplifier response has three main characteristics: its offset or bias (Sect.~\ref{ssec:bias}), its slope and its non linearity (Sect.~\ref{ssec:linearization}). The slope is the gain of  the digital conversion process, given in units of ADU e$^{-1}$. The gain is influenced by several reference voltages and the temperature of the front end electronics. They are measured and provided to ground as numerical values in the housekeeping data, associated with each exposure. 
    
The laboratory characterization provides a model of gain $g$ which depends of the input voltages and temperatures. This model is applied in the pipeline to correct each exposure: 
	\begin{equation}\label{gaineq2} 
	I_{\text{g}} =  I_{\text{b}}/g(T, V);
	\end{equation} 
where $I_{b}$ is the input bias-corrected image, $I_{\text{g}}$ the output gain-corrected image, T and V the housekeeping temperatures and voltages. After correction, the resulting image is in units of photoelectrons, and can be used directly to determine the shot noise. Typical measured values of the gain in the nominal setup are around 0.5 ADU e$^{-1}$.

\subsection{Linearization}
        \label{ssec:linearization}
\begin{figure}\centering
		\includegraphics[width=\linewidth]{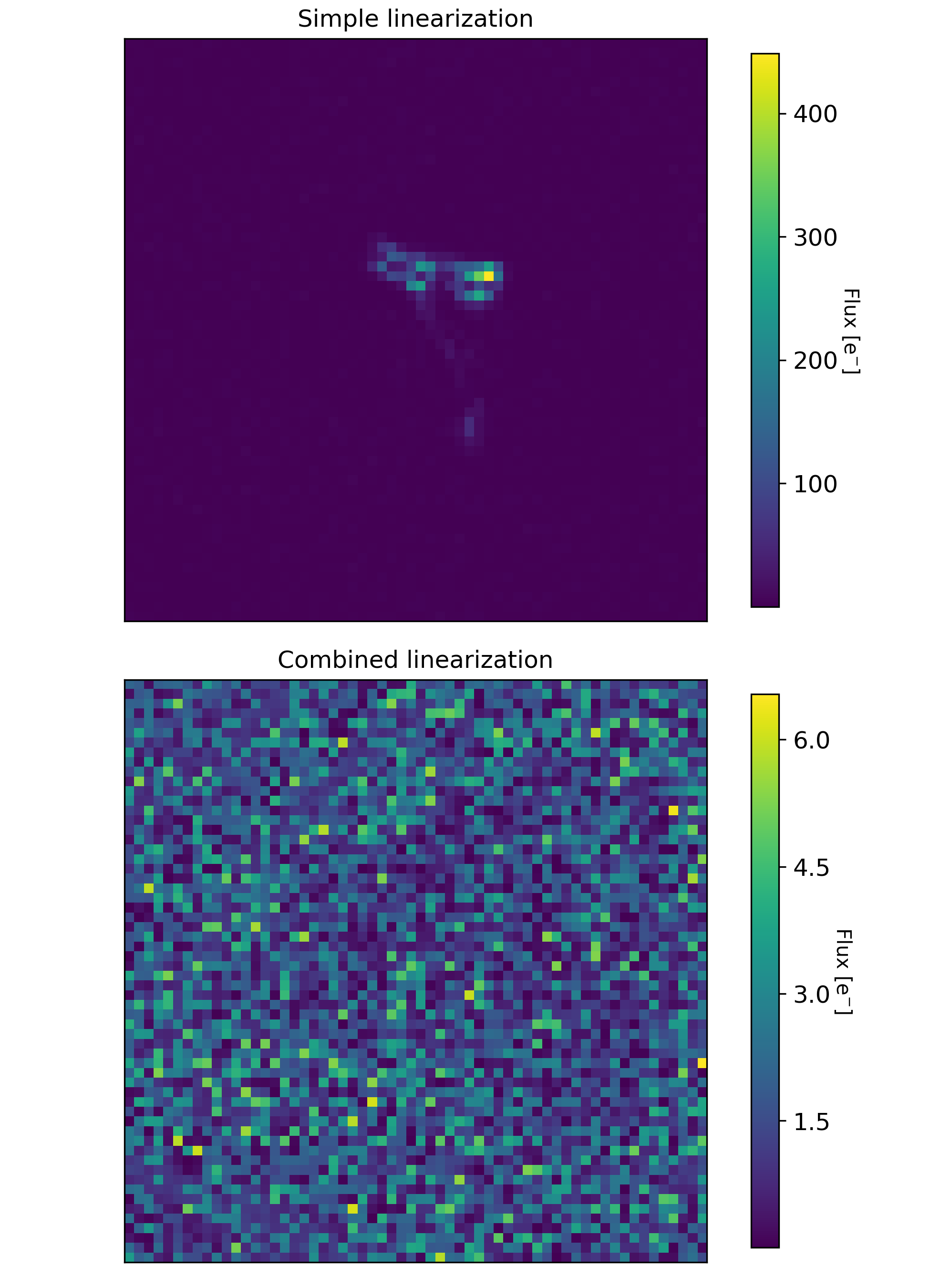}
		\caption{Linearization residuals of a 60x60 pix image of a $V$-mag=9 target built from 6 stacked readouts. Top: from direct application of the linearization correction to the stacked image. Bottom: obtained using the combined algorithm.}
		\label{fig:fig_lincombine}
    \end{figure}

% --------------------------
	The classical linearization is the straightforward application of a correction law determined from laboratory measurements of a constant light beam through a series of increasing exposure times. But such an approach does not work well on stacked images because the correction law itself is not linear. It should therefore be applied to individual readouts prior to stacking. Since the individual readouts are not always downloaded, the pipeline takes advantage of the imagettes to mitigate this limitation. The linearization of imagettes is combined with the linearization of stacked images when necessary. Because the position of the imagette may change at each readout to follow the target's motion, some pixels are not present in all imagettes of a given stacked image. Therefore, the algorithm completes the missing information by properly weighting pixels taken from the stacked image. Figure~\ref{fig:fig_lincombine} shows the efficiency of this technique to restore linearity on the illustrative case of a $V$-mag=9 star whose images are built from 6 stacked readouts. The combined linearized stacked image on the bottom panel shows no imprint of the PSF compared to the classical linearization shown on the top panel where residuals of PSF are clearly visible. It is worth noticing the difference of intensity scale between the two panels.  In conclusion, the linearization is applied to the gain-corrected image $I_g$ to obtain the linearized image $I_L$. This step involves no change of units.

  As an alternative to the on-ground processing, on-board correction of the non-linearity could be performed. This will be investigated during the instrument commissioning, through a series of dedicated tests and, depending on the results, the pipeline could be updated accordingly.

\subsection{Dark current}
    \label{ssec:dark}
	% --------------------------
	
The dark current is accumulated by a given pixel of an image from the beginning of the exposure to its readout. The dark current $d_{\text{e}}$ is monitored using dedicated blind pixels on either side of the CCD: the 32 dark columns which are not exposed to light (see Fig.~\ref{fig:pixmap}).
	
In the default configuration, the readout process starts by quickly transferring the full frame image, including margins, into the blind storage zone in  $t_{z} = 25$ ms. The image is then shifted down line by line during a total time of  $\sim4.63$ s into the serialization register where each pixel is in turn  shifted into the digitizing electronics in a time $t_{x}$ that depends on its $x$-position (column number).
	
Consequently, each pixel has a different dark current time accumulation,  depending on its ($x,y$) position on the CCD. This is described by time map $M(x, y)$ of the same dimensions as the CCD:
	\begin{equation}\label{eq:timemap} 
	M(x,y) = n\cdot (t_{exp} + t_{z} + y \cdot t_{y} + x\cdot t_{x}),
	\end{equation}
where $n$ is the number of stacked images, $t_{exp}$ is the integration time of an individual image, $t_{y}$ the line shift time and $t_{x}$ the column shift time. The dark current estimation is a robust linear regression between dark pixels values and their lifetimes in the time map $M$. The typical dark current is $\sim0.05~ \mathrm{e}^- s^{-1}$, resulting in $\sim1800$~e$^-$ in a typical photometric aperture during one minute exposure. 

To save telemetry, the dark margins are line by line averaged on-board into a single column of 200 mean dark pixels (same y-axis size of the subarray). The time map $M$ is averaged accordingly before the regression. Median and standard deviation are also provided as robust backup and controls in case a cosmic ray would hit the dark columns. 
	
Similarly to the case of the bias correction (Sect.~\ref{ssec:bias}), the dark current correction is separated into a constant and a variable terms to optimize the SNR of the correction. 
The constant term is accurately determined by averaging the dark current over the full observation run. The correction of the variable component is left to the background correction step (see Sect.~\ref{ssec:bkg}). 
	
Finally, the dark current difference between two pixels is corrected by applying a fixed dark frame, $D$. The latter is derived from laboratory measurements. It has to be properly scaled in order to match the actual in-orbit conditions. The dark frame will be updated during the commissioning with exposures with the cover closed. The complete correction of the dark current constant term is then given by:
\begin{equation}\label{eqdark} 
	I_d =  I_{L} - d \cdot M \cdot D\frac{d}{d_{\text{lab}}},
	\end{equation}
where $I_d$ is the dark-corrected image, $I_L$ is the image after linearization in units of electrons, $d$ is the constant in-orbit dark current and $d_{\text{lab}}$ the constant dark determined in laboratory conditions.  

\subsection{Flat field}
    \label{ssec:FF}
	% --------------------------
	
\begin{figure}
		\includegraphics[width=\linewidth]{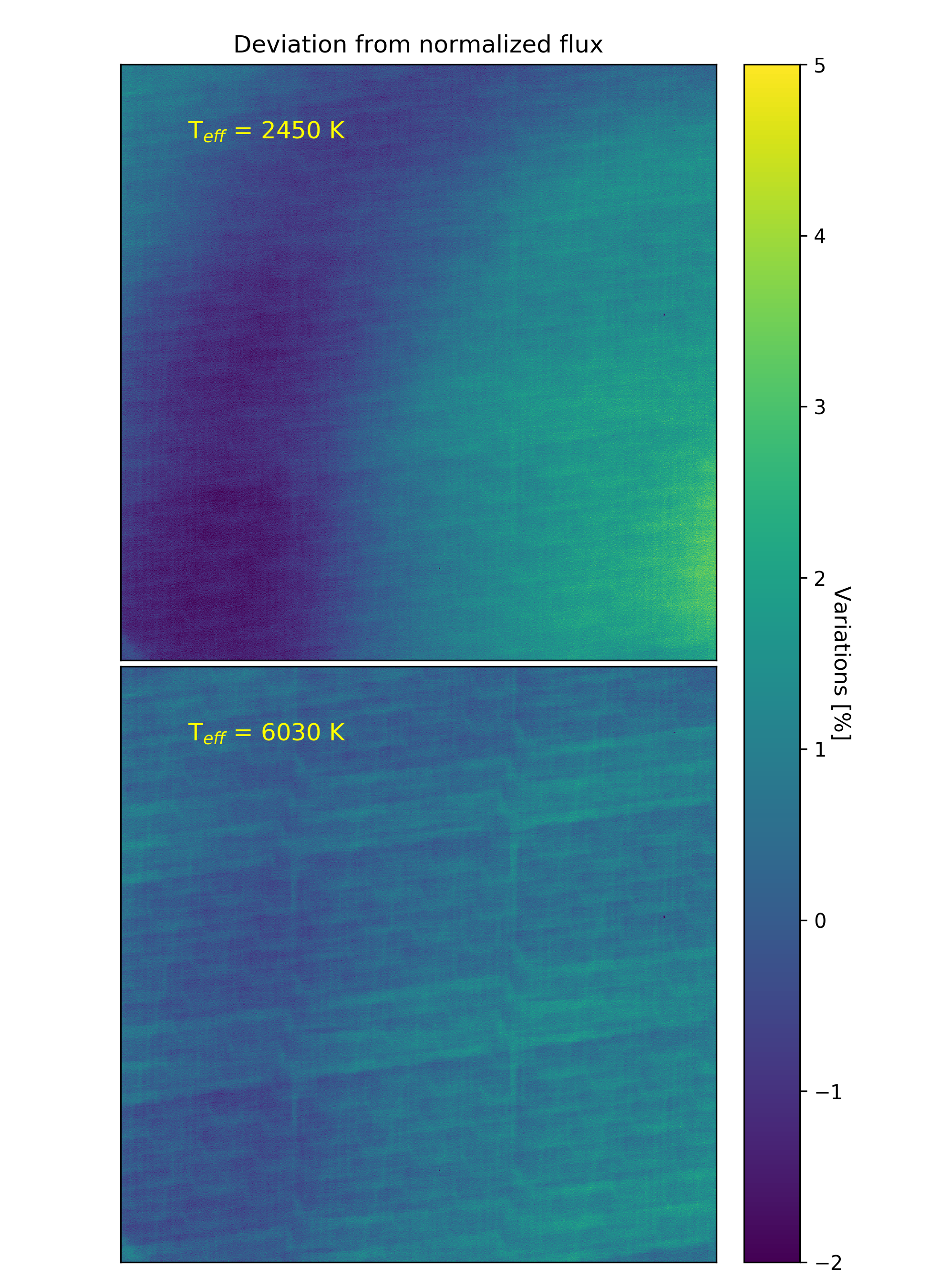}
		\caption{Examples of flat field images derived from monochromatic images corresponding to the spectral energy distribution of a (top) $T_{\text{eff}}=2450~\mathrm{K}$  and a (bottom) $T_{\text{eff}}=6030~\mathrm{K}$ star.} 
		\label{fig:flat_field}
	\end{figure}

\Cheops\ uses a chromatic flat field correction to take into account the PRNU. The dependency of the flat field with the wavelength has been carefully assessed in the laboratory, resulting in a large set of monochromatic images (Deline et al. \textit{in press}). Depending on the wavelength, these measurements show noticeable structures: surface gradients for long wavelengths and strays for short wavelengths as can be seen in Fig.~\ref{fig:flat_field}.

The flat field used for correction is a linear combination of several monochromatic measurements, weighted according to the effective temperature $T_{\text{eff}}$ of the target. The determination of the input $T_{\text{eff}}$ of the target is responsibility of the scientific user or observer of the visit. The \DRP~ will use this value automatically as an input.
A set of mean normalized $T_{\text{eff}}$ indexed flat fields spaced out by $\sim150$~K is available for the correction. Thus, the pipeline uses the flat field image that better matches the target's temperature to perform the correction.

The flat field correction is the last stage of  the calibration module of the \DRP. The calibrated image is in units of photo-electron and passed to the correction stage (Sect.~\ref{sec:cor}).

\section{Correction}
    \label{sec:cor}
	% -----------------------
The correction step aims at correcting individual calibrated frames from environmental effects such as smearing trails, bad pixels, and background and stray light pollution, as detailed in the following subsections. The \textit{pixel-to-sky} step presented as the first box in the correction module in Fig.~\ref{fig:Flowchart} is a general purpose function described in Sect.~\ref{sec:Gen}.

\subsection{Smear correction }
    \label{ssec:smearing}
	% --------------------------
Because there is no shutter, the pixels remain exposed during the readout process. Therefore during the 25~ms of the frame transfer, each charge well collects light from each pixel crossed on its way to the storage area. As a result, vertical trails do appear on the image (top panel of Fig.~\ref{fig:smearcor}). The trails are generated by all stars on the CCD even when located outside the subarray image.

\begin{figure}
		\includegraphics[ width=\linewidth]{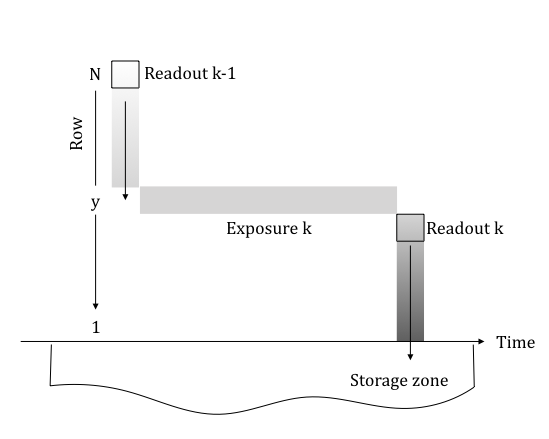}
		\caption{Illustration of the process of charge transfer.}
		\label{fig:smear_cartoon}
	\end{figure}

Figure~\ref{fig:smear_cartoon} illustrates this effect in the case where the individual exposures are not stacked.  At the end of exposure $k-1$ an empty charge well is created on the top of the CCD and will reach its integration position $y$ after crossing the upper pixels $N$ down to $y+1$ and collecting a fraction of their flux. When the exposure $k$ begins, this pixel does thus already contain part of its future smear. At readout $k$ it sweeps down trough the rest of the CCD, but across a slightly different image because of the motion of the field. 

As a result, the smear flux $f_k(y)$ collected in pixel $y$ of the image $k$ is:
	\begin{equation}\label{eq:eqsmear} 
	f_k(y) = \sum_{i=y+1}^N s_{k-1}(i) + \sum_{i=1}^{y-1} s_k(i),
	\end{equation} 
where  $s_k(i)$ is the flux collected when the charge well passes under photo-site $i$ during readout $k$.  The first and second terms correspond to the contribution of the column above and below the pixel respectively. The smear problem is thus to estimate the contributions of the various photo-sites crossed by a given pixel of the image. 
	
The basic approach would be to derive the contributions $s(i)$ from the image itself as proposed by \cite{Powell1999} who subtracts the summed column of the image properly scaled by the transfer time, or \cite{Iglesias2015} which adapts that principle to varying illumination during the exposure. But, this approach does not apply to \Cheops\ since only a part of the CCD is downloaded and because the image continuously varies over time: between two consecutive 1 min exposures the image is rotated by 3.6$^{\circ}$ and undergoes a different pointing jitter.
	
To estimate the smear, a set of overscan pixels is then available. Overscans take form of 6 rows of virtual pixels on the top of the image (see Fig.~\ref{fig:pixmap}). An overscan is not a silicon pixel but an extra clocking at readout time that generates an empty well which immediately crosses the whole CCD following the image. Therefore, the overscans only contain the smear flux. 

The contribution $s_k(i)$ defined at Eq.~\ref{eq:eqsmear} can be estimated from overscans by:
	\begin{equation}\label{eq:eqovs}
	s_k(i) = \omega_k,
	\end{equation}
where $\omega_k$ is $1/N^{\text{th}}$ of the average overscan row at exposure $k$. The smear Eq.~\ref{eq:eqsmear} then resumes to: 
	\begin{equation}\label{eq:smearrow} 
	f_k(y) = (N-y)\omega_{k-1}+(y-1)\omega_k.
	\end{equation}
The correction consists on subtracting the estimated smear flux from the image. Figure~\ref{fig:smearcor} shows an isolated star of $V$-mag=9 before and after correction. The bright target generates large smear trails (top panel) due to an important number of stacked readouts. In the image on the bottom, the correction has been applied \citep[e.g.,][]{Jenkins2010b, Rauer2014}.  

Although the correction looks fine in the images, it causes a significant increase of noise in the light curve. Due to briefness of the 25~ms transfer time, only few electrons are collected in the overscans and thus cause an important shot noise. That noise is amplified by the large area of the photometric aperture similarly to the bias correction (Sect.~\ref{ssec:bias}). 
 
	\begin{figure}
    \centering
        \includegraphics[scale=0.75]{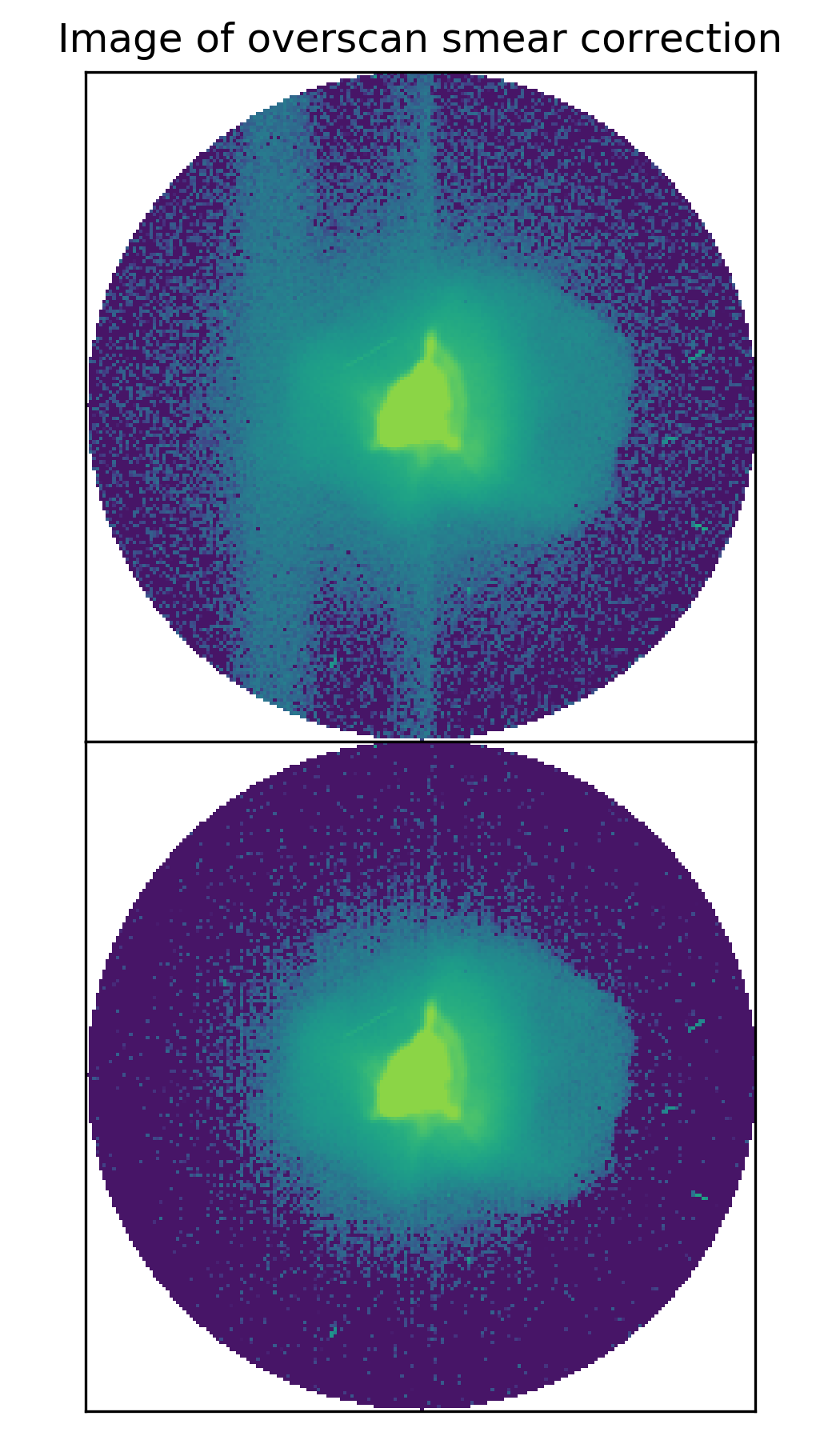} 
		\caption{Top: a simulated exposure of a $V$-mag=9 target and one external contaminant. Bottom: the same exposure after smearing correction. The color-scale has been adapted for better visualization.}
        \label{fig:smearcor}
	\end{figure}

%The smear estimate $\omega_k$ assumes a uniform column, but fortunately, this constraint can be relaxed when the image is still or when the flux is dominated by the target as others transit methods experiments. 
The smear estimate $\omega_k$ from the overscans assumes an uniform column or trail. This assumption holds when the observed image is static or when the smear flux is dominated by the target as it was the case in others missions of photometric observations of transits such as CoRoT and Kepler.

\begin{figure}
    \centering
        \includegraphics[ scale=0.75]{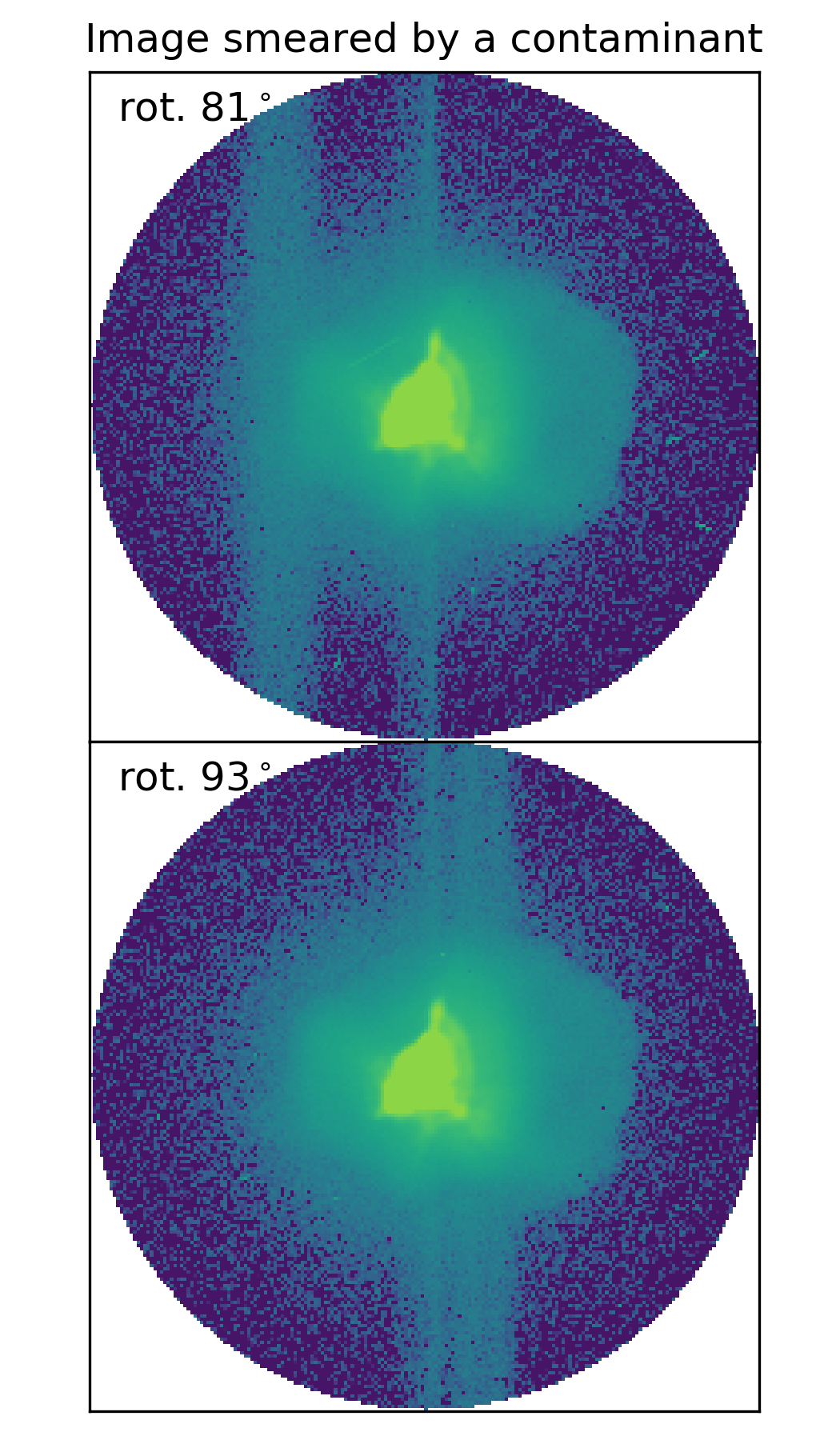} 
		\caption{Smear trail, on logarithmic scale, for 2 different roll angles of a contaminant star of  $V$-mag=7 rotating around a $V$-mag=9 target outside the subarray. Each exposure is labelled with its respective roll angle.}
        \label{fig:smearext}
	\end{figure}

Figure~\ref{fig:smearext} shows that this is not necessarily true for \Cheops. Indeed, the trails of the others stars rotating around the target can overlay the target, even when located far outside of the downloaded region of the CCD. Additionally a star which is present at exposure $k-1$ below a given pixel will leave a trace in the corresponding overscan $\omega_{k-1}$ used for correction, but might have rotated away at exposure $k$ and thus never been crossed by the pixel.

\begin{figure}
		\includegraphics[ width=\linewidth]{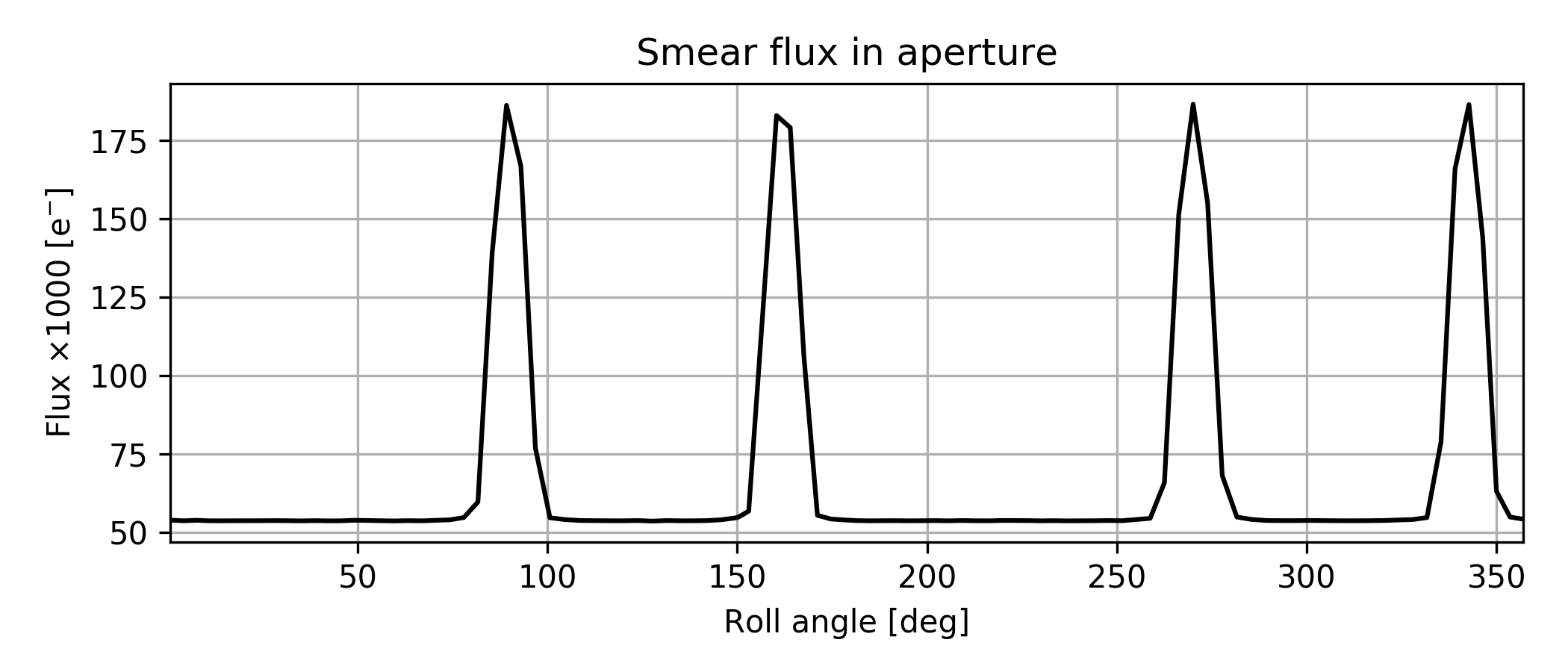}
		\caption{Example of the smear flux in the aperture from the target and external stars as a function of the roll angle.}
		\label{fig:selfsmear}
	\end{figure}

The solution comes from Fig.~\ref{fig:selfsmear} which shows the light curve of the isolated smear pattern obtained by simulation. The peaks of the curve originate from the crossing trail of the an external star of Fig.~\ref{fig:smearext} rotating outside the subarray. %The number and brightness of the stars have been exaggerated for testing purposes. 

It is important to note that the flux outside the peaks, which originate only from the trail of the target itself (hereafter called self-smear) is nearly constant (e.g. the flat bottom of the curve in Fig. \ref{fig:selfsmear}). That comes from the fact that the photometric aperture follows the motion of the target and consequently it \textit{sees} a static pattern. Therefore it is not necessary to correct the self-smear which is left as is to avoid introducing noise. 

Only the peaks above a certain threshold are corrected, using simulated images (Sect.~\ref{ssec:simu}) to determine the concerned exposures. The threshold is chosen to ensure that the noise introduced by the overscan based correction will be smaller than the disturbing smear signal.  

\subsection{Bad pixels}
    \label{ssec:bad}
    
	The \textit{bad pixels module} detects and corrects for cosmic ray hits during the observation as well as for pixels with temporary or permanent abnormal response. Currently, the types of \textit{bad pixels} considered by the \DRP\ are:
	
	\begin{itemize}
		\item Cosmic rays. When high energy particles impacts the CCD they cause positive outliers in a pixel during a single exposure. These cosmic rays (CR hereafter) can affect one or several connected pixels, as well as dark and overscan CCD margins. The CR hits occur mainly during the SAA crossings that are not down-linked, but also spuriously outside the SAA.  
        \item Hot and dead pixels are permanent damaged pixels that suffer abnormally high or low flux response, respectively.     
		 \item Random Telegraphic pixels are unstable pixels whose state randomly flips between a normal behavior and an arbitrary high response, or just are affected by a high level of noise. Caused by irrecoverable radiation damages, the total number of telegraphic pixels is expected to increase during mission lifetime.  
	\end{itemize}
	
 The CCD will be regularly monitored and an updated list of bad pixels will be issued and serve as an entry for \DRP. The pipeline then notifies after each observation its own detections from the signal in the subarray window. The location of the subarray will be chosen to avoid hot/dead pixel in the aperture.

 \begin{figure}
		\includegraphics[width=\linewidth]{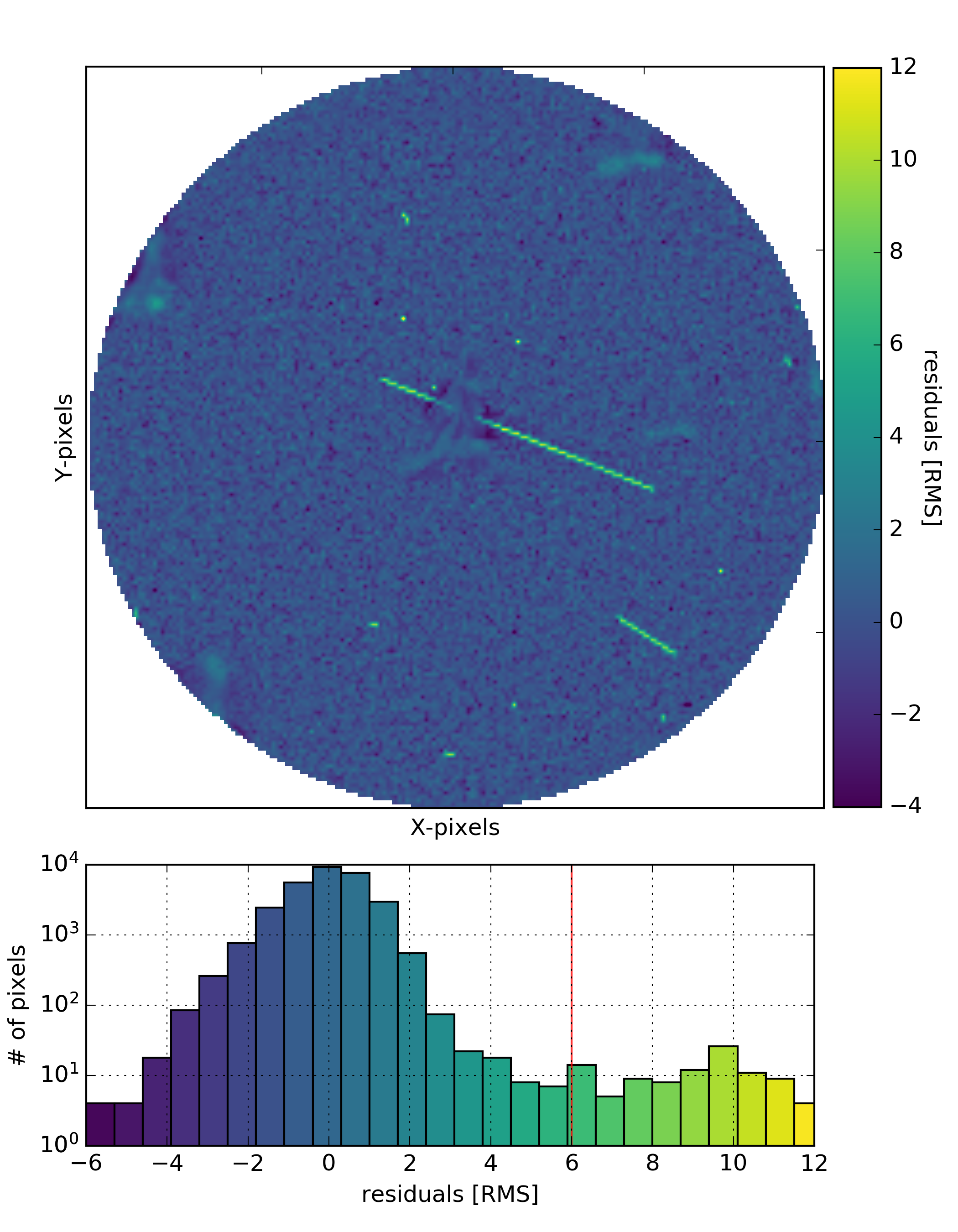}
		\caption{Top: example of a residuals image used for cosmic rays detection. Bottom: residuals distribution of the image. The vertical line represents the detection threshold.}
		\label{fig:CR_res_sigma}
	\end{figure}	
    
      \begin{figure}
		\includegraphics[width=\linewidth]{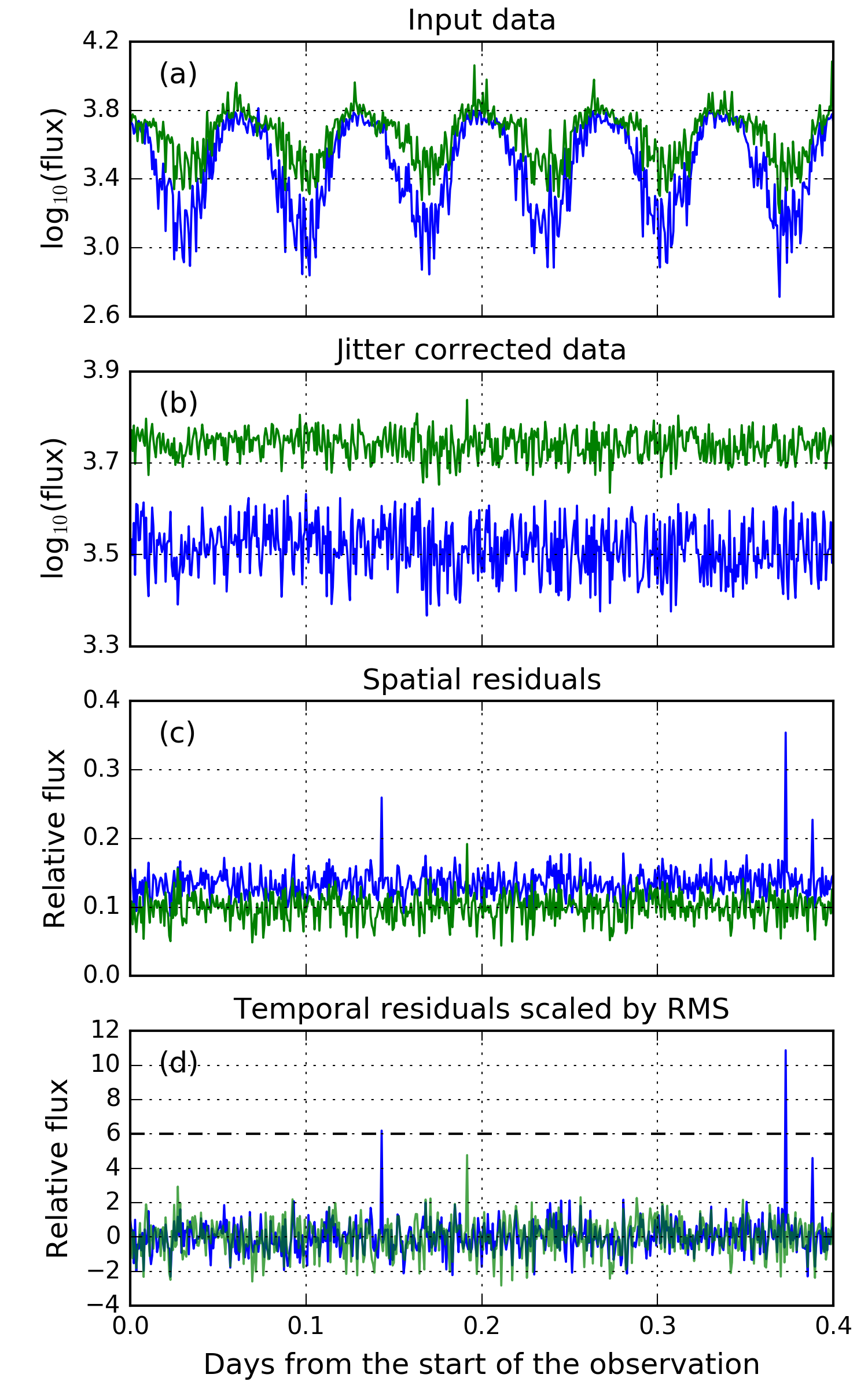}
		\caption{Evolution of the light curves of two pixels close to target along the CR process. {\it Panel (a)}: initial flux. {\it Panel (b)}: flux after re-centering by the opposite depointing direction. {\it Panel (c):} Spatial residuals. {\it Panel (d)}: residuals normalized unitary RMS. The horizontal dashed line represents the current 6-$\sigma$ threshold used to flag the cosmic ray hits.}
		\label{fig:CR_lcs}
	\end{figure}  	
	
Simple approaches like sigma clipping to search for outliers in pixel flux time series is not relevant for \Cheops\ because of the specific features of its data such as: i) the noise is not stationary due the permanent rotation of the image and ii) at pixel level the noise is dominated by the jitter noise, especially for the peaks of the PSF near the center of the target. To reduce the temporal variability of individual pixels due to the jitter, the bad pixels detection module begins by re-centering images and imagettes, shifting them in the opposite direction of the depointing. Then, to remove flux variations caused by the rotation of the images and the target's intrinsic variability affecting close pixels in the same way, the detection of bad pixels operates on the residuals. A residual $r$ is the relative variation of a pixel compared to its neighbours:
	\begin{equation}\label{eq:bpresiduals} 
	r = \log(\frac{f}{f \ast k}),
	\end{equation}
where $f$ is the image (resp. imagette) and $k$ an unitary smoothing kernel of size $10\times10$ (resp. $6\times6$) pixels. The sign $*$ accounts for convolution function. The advantage of using residuals is that each type of bad pixel has an specific footprint in these $r$ images.  

Special attention is brought to CR which are difficult to detect when embedded into the target main flux. For this reason, the detection is also performed in the imagettes where the same CR flux has a better contrast with respect to the reduced unstacked target flux. Both detections maps are then merged, taking into account the fact that the target's position follows the depointing. 

 As temporal outliers, CR are detected by sigma-clipping the residuals. The adopted threshold is adjusted to represent the best compromise between the number of detections and the number of false positives, avoiding the correction of false events. The thresholds are derived from a set of simulations over a wide range of target brightness  ($6<V$-mag$<12$) and exposure times (see Sect.~\ref{sec:performance}). Figure~\ref{fig:CR_res_sigma} shows an example of a residuals image used for the CR detection. One long trail of a cosmic ray crossing the target's PSF is detected in the upper image, except for the pixels inside the peaks of the PSF. The CR energy is indeed not large enough to stand out among the PSF pixels flux. The 6$-\sigma$ detection threshold is represented by the vertical line in the lower histogram of residuals. All pixels above this threshold are flagged as CR. The evolution of the light curve of two pixels through the CR detection module is shown in Fig.~\ref{fig:CR_lcs}.

Once the bad pixel detection is performed, the \DRP\ proceeds to the correction of the pixels hit by CR. This correction is done with a 2D cubic interpolation of neighbor pixels using the \texttt{Python} routine \texttt{interpolate.griddata} of the \texttt{Scipy} library. 
    
\begin{figure}
		\includegraphics[ width=\linewidth]{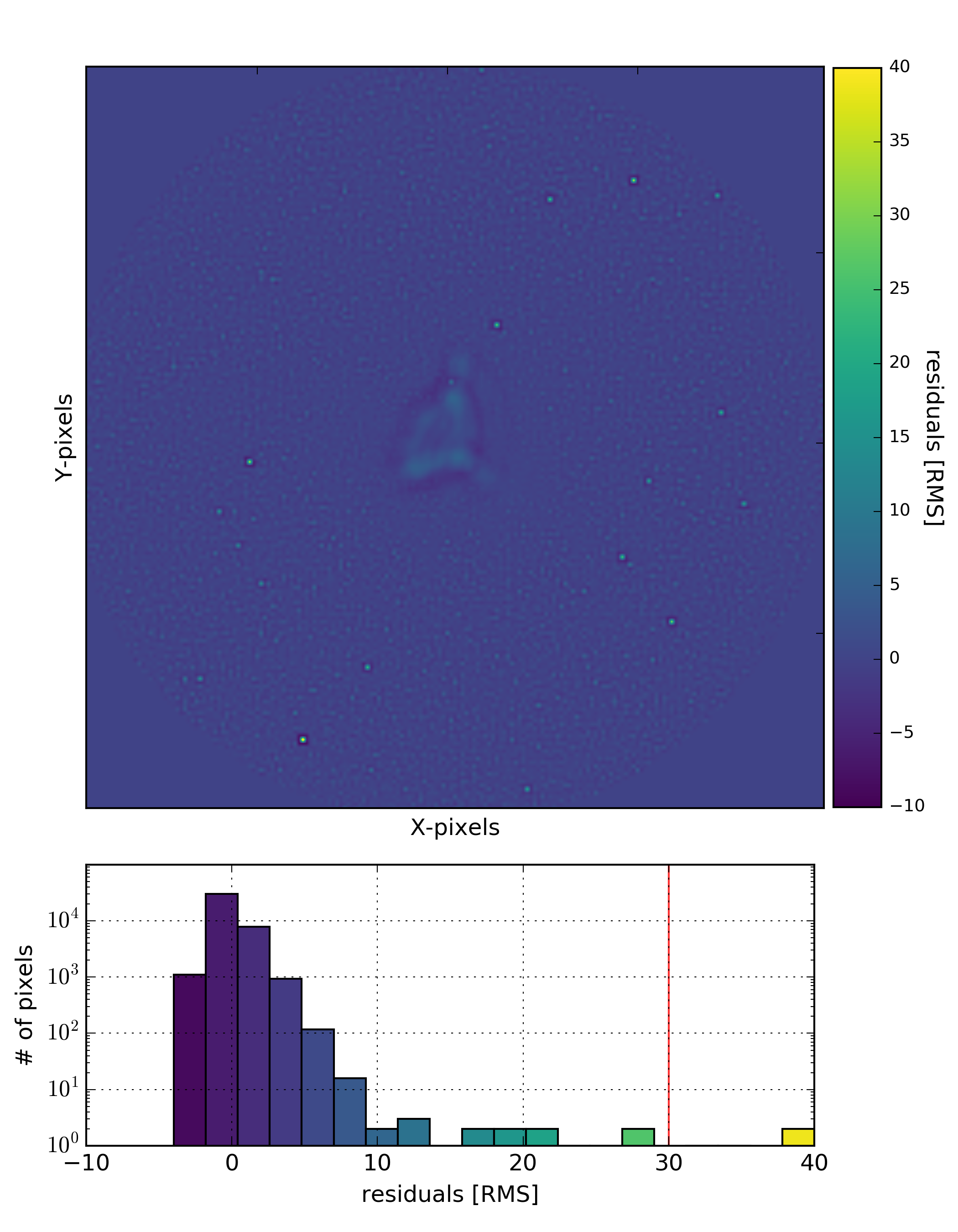}
		\caption{Top: Temporal mean of residuals normalized by the spatial MAD (median absolute variation) as an 200x200 pix image. Bottom: histogram of the distribution of the residuals mean values where the vertical line marks the detection threshold of hot pixels.}
		\label{fig:hotmap}
	\end{figure}

Hot (dead) pixels are positive (negative) spatial outliers imprinted, in this case, on the temporal average of the residual images. They are detected via a spatial sigma clipping. A threshold as high as 30-$\sigma$ is necessary to avoid flagging pixels influenced by the peaks of the target's PSF. No centering is applied before hot, dead and telegraphic pixels detection. An example of a temporally averaged residuals is shown in Fig.~\ref{fig:hotmap} for a target of $V$-mag=6 . The signatures of the PSF are clearly visible at the center of the image. Hot pixels appears as strong positive values on this map while dead pixels as negative ones. 
 
\begin{figure}
		\includegraphics[ width=\linewidth]{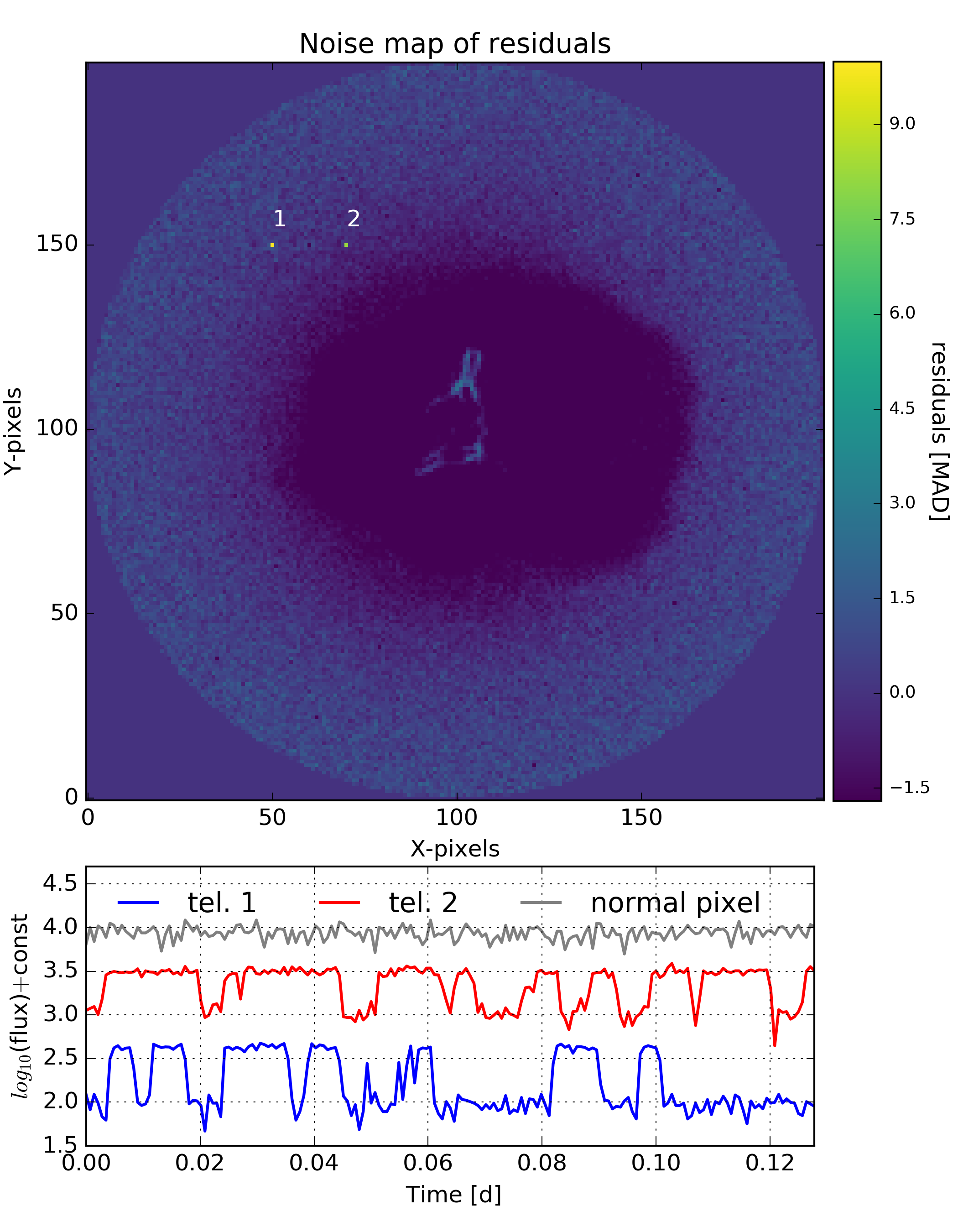}
		\caption{Top: temporal noise of residuals normalized by the spatial MAD. Bottom: Light curves of the two flagged telegraphic pixels (green and blue lines) and one normal pixel (gray line).}
		\label{fig:variationmap}
	\end{figure}

Finally, the telegraphic pixels are detected as noisy pixels in the map of residuals variation over time. On the contrary, residuals of an ordinary pixel shows small variations over time. Nevertheless, a detection threshold of 7-$\sigma$ is used to avoid false detections in the PSF peaks. An example of a noise map is shown in Fig.~\ref{fig:variationmap}. The effect of the jitter on the target is evident at the center.

 Finally, the Bad Pixel module outputs the corrected image cube and the 2D map of the bad pixel location.  

    \subsection{Background}
    \label{ssec:bkg}
	The zodiacal light, non resolved background objects, stray light from Earth and Moon inject a non constant flux offset over the CCD. This background flux depends primarily on the orbital phase and on the pointing of the telescope. In particular for \Cheops, the background correction module plays an important role because of the satellite's proximity to the Earth. The classical approach of background estimation based on selected background windows gives poor results for \Cheops\ since the displacement of the stars due to the rotation of the image obliges to move them continuously, probing thus not all the time the same pixels and flux distribution. These changes translate into discontinuities in the estimated background time series which is ultimately imprinted in the light curve of the target by the correction.
	
\begin{figure}
		\includegraphics[width=\linewidth]{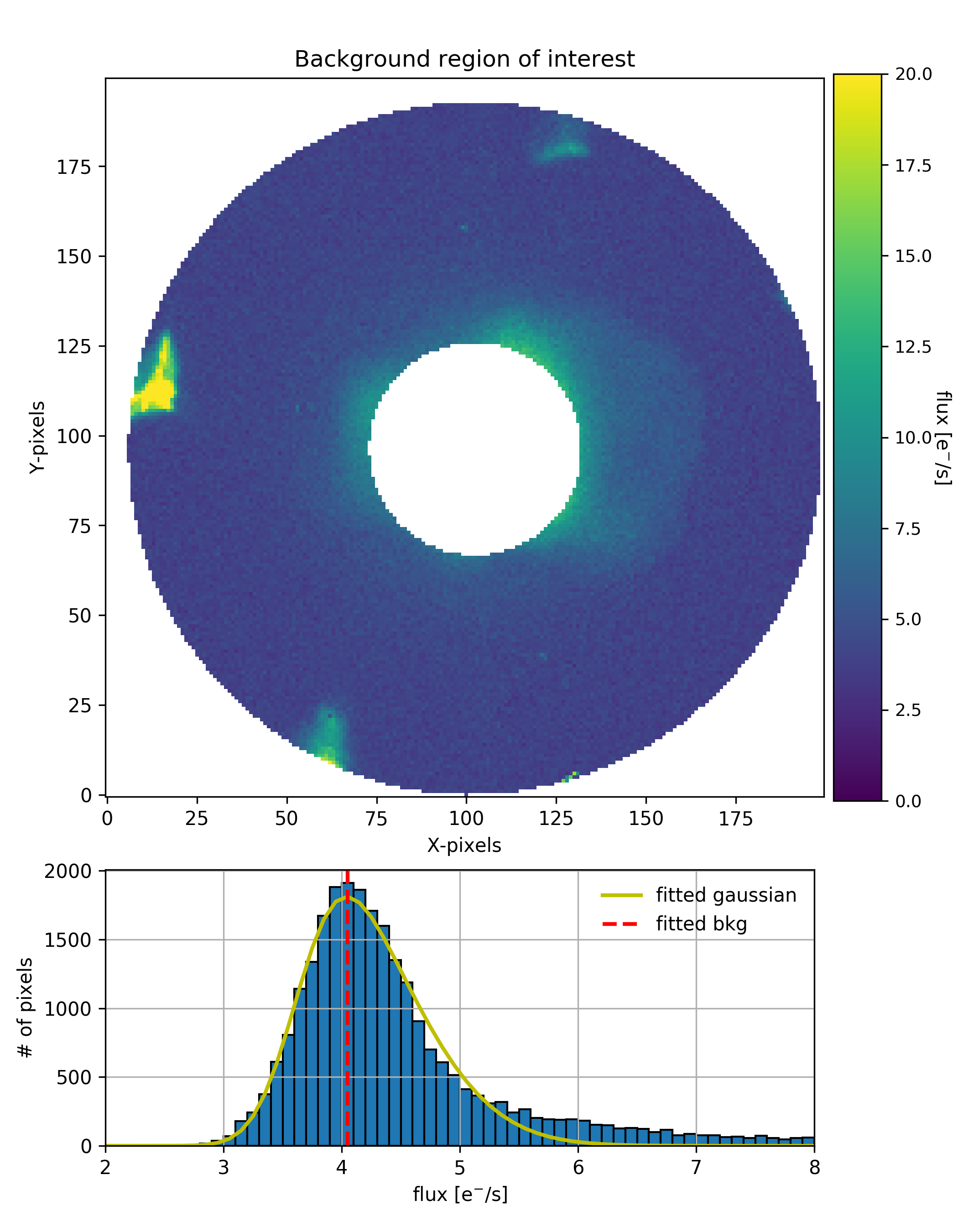}
		\caption{Background estimation from one image. Top: masked region excluding the target. Bottom: histogram of the pixels in the background region after clipping extreme values shown together with the fitted Gaussian function (orange) and the adopted background level (red dashed vertical line). }
		\label{fig:bkgmask}
	\end{figure}

\begin{figure}
	   	\includegraphics[width=\linewidth]{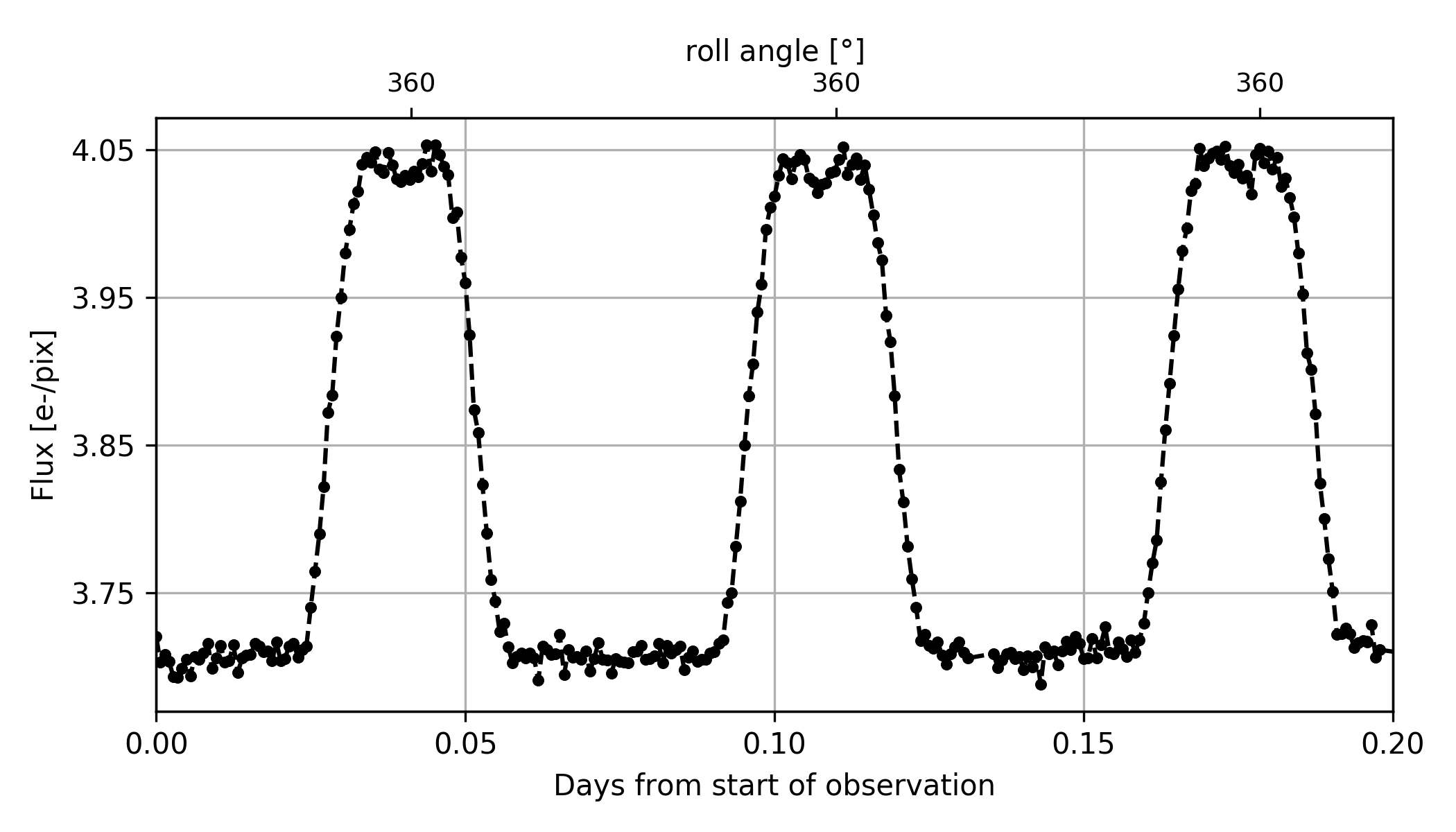}
		\caption{Example of a background curve of a 5-hour observation under typical observing conditions. The correlation with the roll angle (top axis) is evident.}
		\label{fig:bkgcurve}
	\end{figure}

	Instead, \DRP\ uses a histogram based method which is insensitive to the rotation of the field and maximizes the total background sampled flux. For this purpose, a large circular mask that excludes the central target is applied to each frame. This mask follows the depointing so that the probed region is always the same. An histogram is drawn from all pixels included in the background mask. The upper bound of the histogram is restricted to the admissible  background level in order to exclude contaminating stars as well as the tails of the target's PSF. Then, the mode of a fitted skewed Gaussian is taken as the background value and subtracted from the image. Figure~\ref{fig:bkgmask} shows an example of the mask (top panel), its respective histogram (bottom panel) and the resulting background estimation (dashed line). The background time series of a typical observation with a few faint background stars is shown in Fig.~\ref{fig:bkgcurve}. There is a clear correlation between the roll angle variation, and therefore the stray light level, and the background flux. For each visit, the background time series is delivered by the \DRP\ and the corrected images are used as the starting point of the photometry extraction. 
  	
	\section{Photometry}
    \label{ssec:phot}
	% -----------------------
	
	After the data has been fully calibrated and corrected, the \DRP\ performs an aperture photometry to deliver the final light curve. The aperture is a circular binary mask that follows the target's displacements. The circular shape respects the intrinsic symmetry of the rotating experiment. To avoid sharp edges like a binary step-like contour, the border is weighted in relation with the pixels fraction covered by the mask.
	
	To avoid area changes when the border shifts from a subpixel quantity, for a particular radius, only one disk template is computed using a null depointing and then applied to all depointings of the whole time series by using an antialiased shifting algorithm that strictly preserves the mask surface. Apertures of non-constant area would introduce artificial photometric noise in the light curves. 
	
	In fact, \DRP\ provides four light curves each measured trough a different aperture. The three first radii are pre-defined: 26, 33 (default aperture) and 39 pix, while the fourth radius is optimized for each visit (optimal aperture). 
	
\begin{figure}
	   	\includegraphics[width=.97\linewidth]{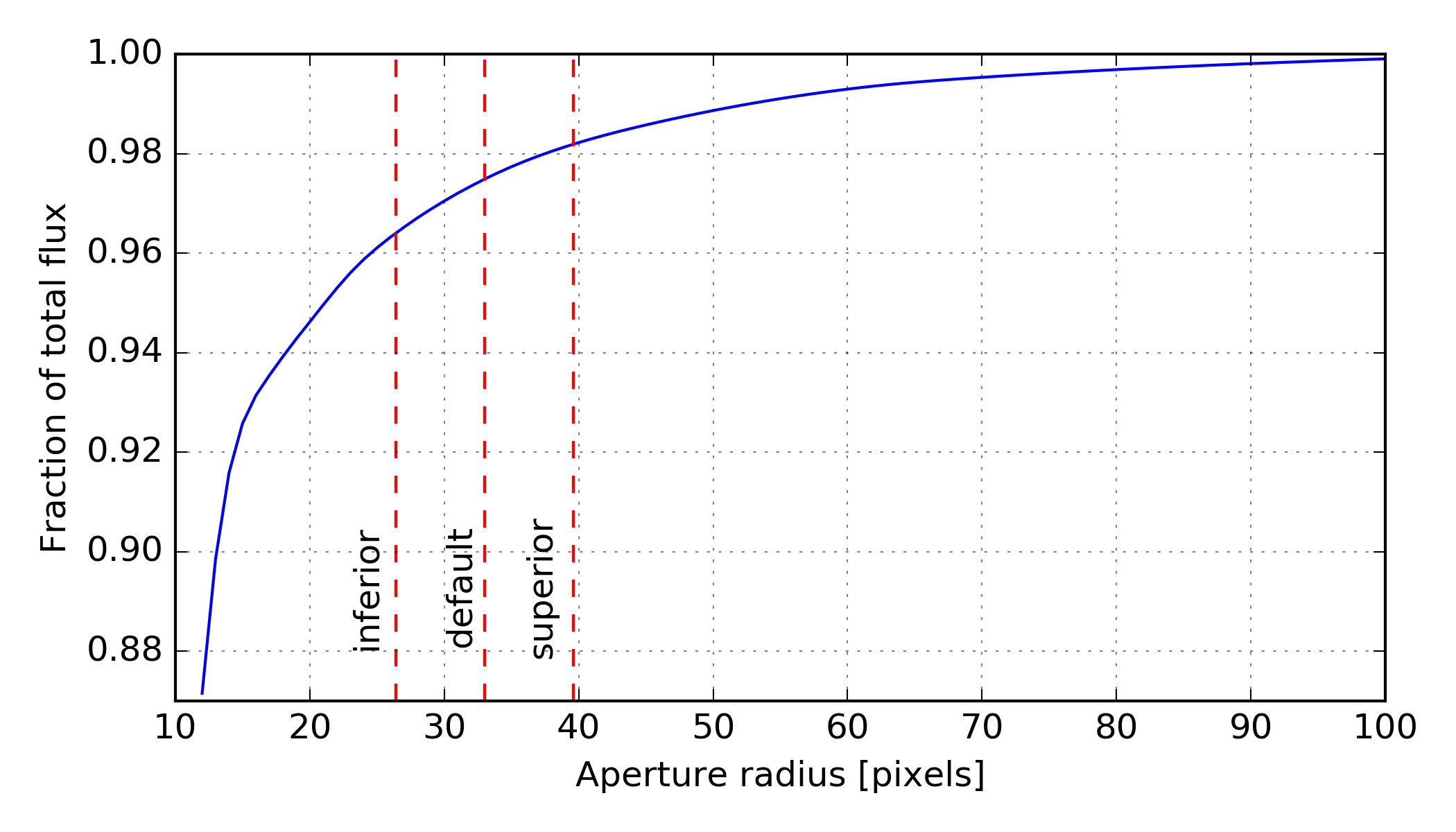}
		\caption{Photometric growth curve of \Cheops\ PSF. The vertical dashed lines represent the radii of the three pre-defined apertures used by the \DRP.}
		\label{fig:growthcurve}
	\end{figure}

	The default aperture (33 pix) encompasses $97.5\%$ of the PSF flux. %It was adopted as a reference for dimensionning \Cheops\ since being satisfactory in most common cases. 
	The two other pre-defined apertures are lower (80\%) and upper bounds (120\%) of the default radius and used as controls. Figure~\ref{fig:growthcurve} shows the flux of the PSF encompassed by each of the predetermined apertures. 
	
	The light curve $f$ is simply the sum of the pixels inside the aperture, and weighted by the mask $m$ depointed by $(\delta x, \delta y)$: 
	\begin{equation}\label{eq:flux}
	f = \sum_{pix} m(\delta x, \delta y) \cdot p,
	\end{equation}
    with $p$ being the concerned pixel. 

	The optimal aperture is optimized for the visit. For instance, bright targets deserve a larger mask as their flux dominate further out from the center over the background and the readout noise. On the other hand dense star fields require a smaller aperture to better exclude contaminating stars. 
	
\begin{figure}
	   	\includegraphics[width=.95\linewidth]{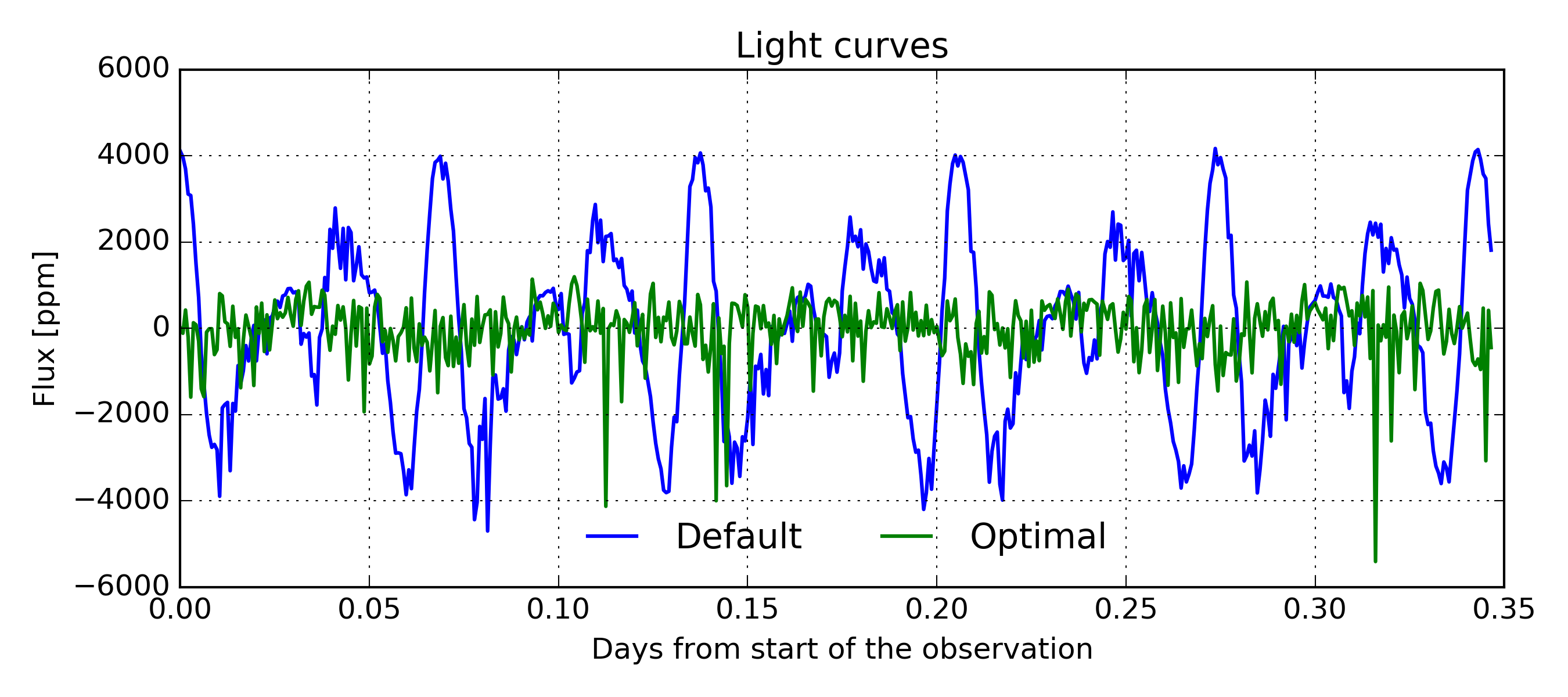}
		\caption{Light curves of the default radius aperture (blue) and optimal aperture (green) for a $V$-mag=6 target with a $V$-mag=9 background star located at $\sim$30 pix distance.}
		\label{fig:lc_def_vs_optimal}
	\end{figure}

\begin{figure}
	   	\includegraphics[scale=0.6]{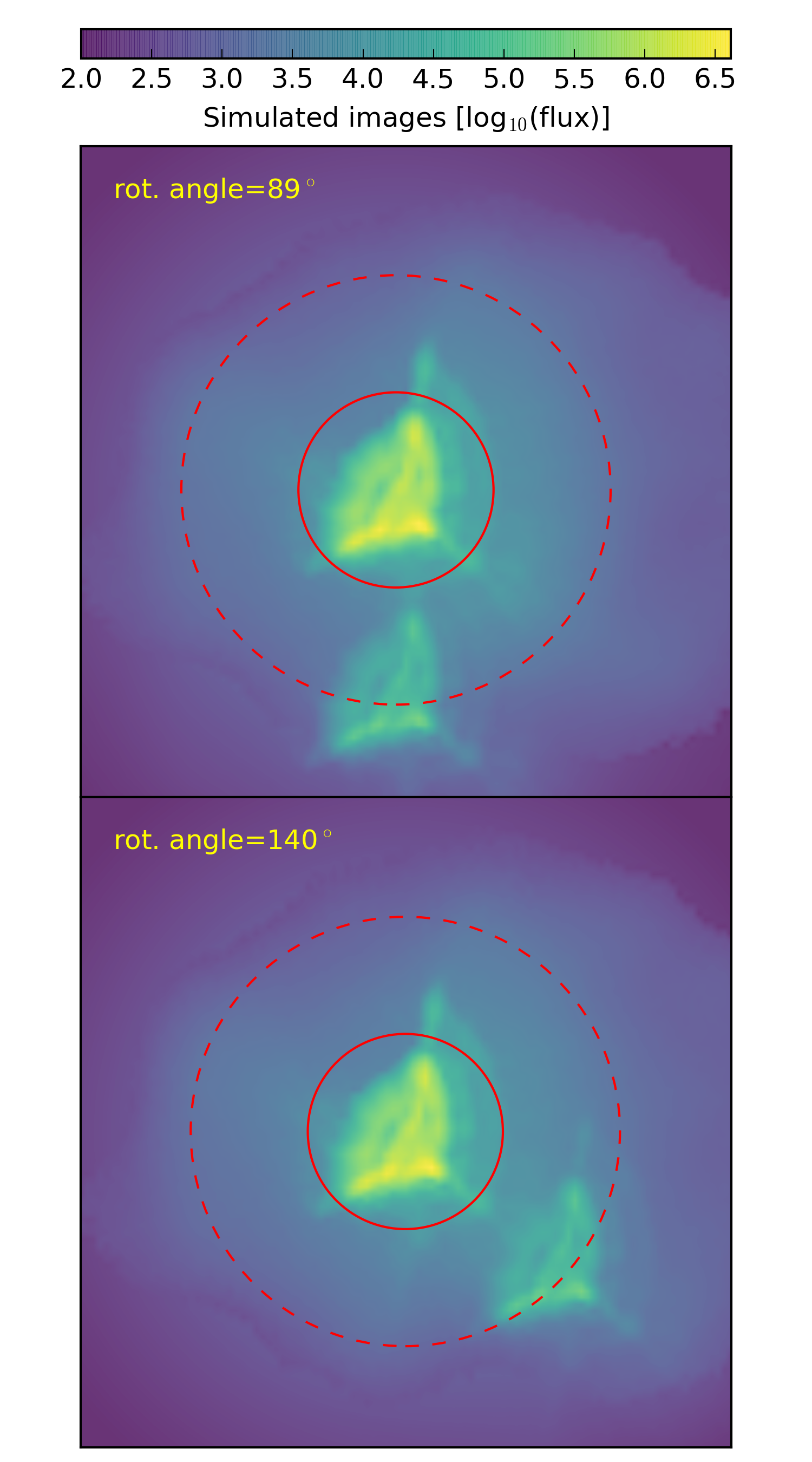}
		\caption{Examples of two simulated images of an observed field composed of one $V$-mag=6 target star and one $V$-mag=9 background star located at $\sim$30 pix distance. The optimal and default aperture for photometry are represented by the solid and dashed red circles, respectively. Each image is labelled with their respective roll angle.}
		\label{fig:stamps}
	\end{figure}

	The optimal aperture corresponds to the radius that minimizes the noise to signal ratio:
	\begin{equation}\label{eq:sn}
	\text{NSR} =\frac{\sqrt{f+c+\sigma_\text{c}^2+\sigma_\text{ron}^2}}{f},
	\end{equation}
where the numerator lists all the considered noise sources. The components $f$ and $c$ respectively accounts for the target and contamination shot noise inside the tested aperture. They are computed from image simulations (Sect.\ref{ssec:simu}). The noise $\sigma_\text{c}$ is the contamination variation caused by the ingress-egress of the contaminants, whose irregular PSF enters and exits the aperture mask along the rotation. The noise $\sigma_\text{ron}^2 = n_{\text{pix}} \cdot n_{\text{stack}}\cdot \text{ron}^2$ is the readout noise estimated in Sect.~\ref{ssec:bias}, and transformed to electrons using the gain, of the $n_{\text{pix}}$ pixels of the mask for an image composed of $n_{\text{stack}}$ stacked readouts. 
	Figure~\ref{fig:lc_def_vs_optimal} shows the photometric improvement when using the optimal aperture if a $V$-mag=9 contaminant is distant from the $V$-mag=6 target by only 30 pixels. The default light curve is clearly degraded by the variable overlapping of the contaminant (Fig.~\ref{fig:stamps}). These flux variations are no longer present when applying the optimal circular mask which in this case was set by the optimization method with a radius of 15 pixels. 
	
	Finally, besides the four light curves, the pipeline delivers as products complementary correction values that could help the user to perform a deeper analysis of the data. Among these products are the dark current, background and contamination light curves. 

	\subsection{Image simulations}
    \label{ssec:simu}
 
    The pipeline builds up simulated images of the whole visit because it needs to estimate smear trails (Sect.~\ref{ssec:smearing}) and contamination from the resolved nearby stars. The \DRP\ internal simulator starts by making use of the World Coordinate System (WCS) of each exposure (see Sect.~\ref{ssec:skycoord}) jointly with the sky coordinates and the CHEOPS magnitudes of the stars extracted from an input catalogue. This catalogue is built for each observation by extracting from the Gaia DR2 catalogue \citep{gaiadr2} the sky coordinates, the CHEOPS magnitude (obtained from the V- or G-band conversion) and the $T_{eff}$ of each star in the field of view, and it is provided to the \DRP\ as an input file associated to each observation. The internal simulator then uses this information to spread a reference PSF over the CCD coordinates of the stars with the flux scaled according to their CHEOPS magnitude, resulting in the expected simulated data set. The reference PSF comes from laboratory measurements during pre-launch instrument characterization. It will be later on replaced by the flight PSF derived from commissioning phase.
    
\begin{figure}
		\includegraphics[scale=0.6]{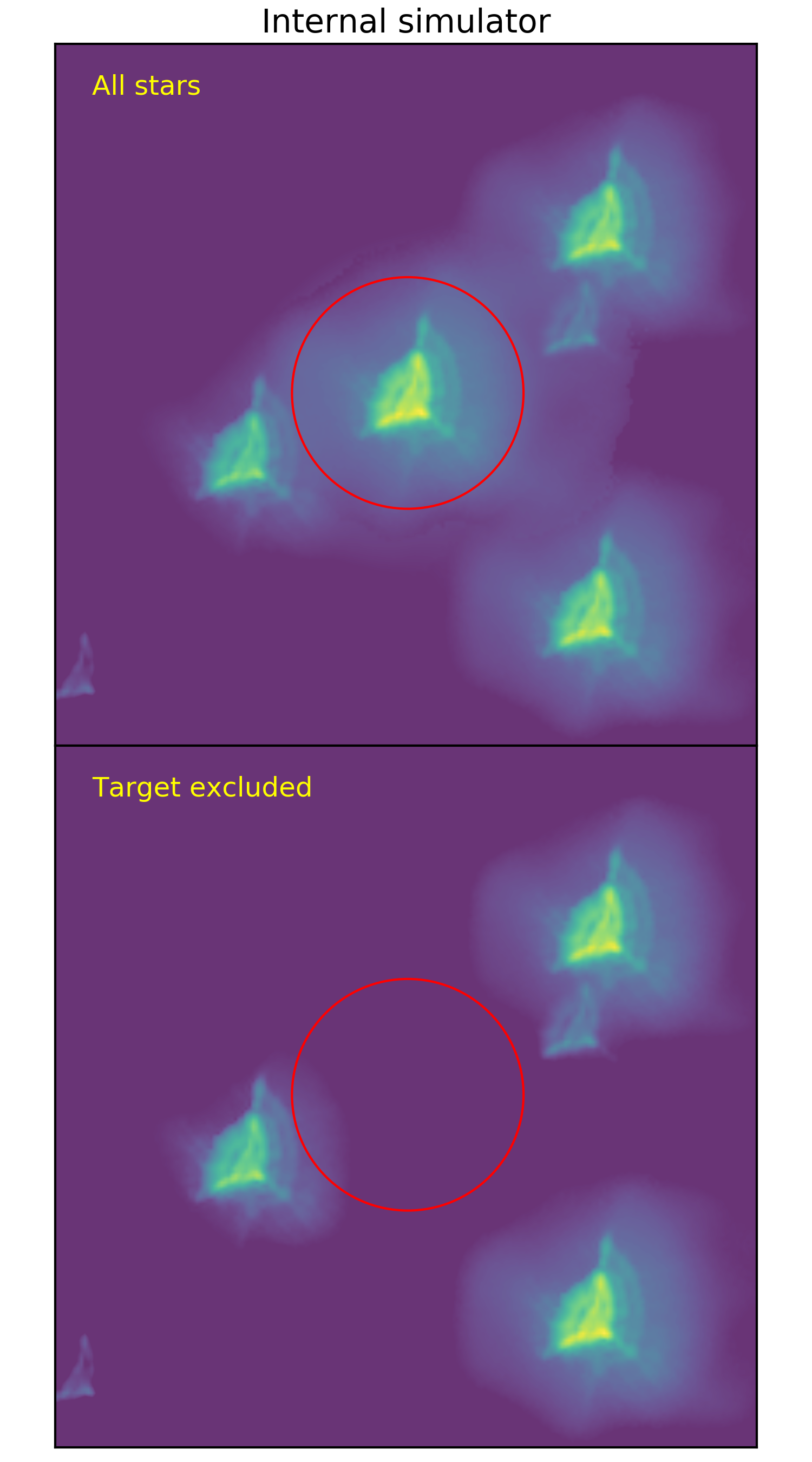}
		\caption{Simulated image of the FoV including the target and all background stars (top) and all the stars but the target (bottom). The red circle represents the photometric aperture.}
		\label{fig:simulations}
	\end{figure}

    A double simulation is first built up: one with only the target in the field and the other with the resolved contaminants only. This pair is used to estimate the effect of the contaminant stars in the photometry (Fig.~\ref{fig:simulations}) and to compute the optimal aperture for the photometry (Sect.~\ref{ssec:phot}). In the figure the red circle represents the location of the photometric aperture that is used to compute the values $f$, $c$ and $\sigma_{\text{c}}$ in Equation~\ref{eq:sn}.
    
    A second simulation over the whole CCD height which includes the non downloaded portion above the image is necessary to model the smear trails since any contaminant crossing that region let a trace in the smear trails. As the smear trails extend along the y-axis, the computation of their simulation is optimized by collapsing both PSF and star coordinates in the single spatial dimension of the y-axis. Here, the possible change of position during the exposure, which will produce a small dilution of the signal on the x-axis is not taken into account at the moment.    
   
    \section{General purpose modules}
    \label{sec:Gen}
	% --------------------------
	
	\subsection{Events flagging}
    \label{ssec:event}
	% --------------------------
	Due to the low orbit, a significant fraction of measurements are lost due to the proximity of the Earth to the line of sight and the crossing of the SAA. The event-flagging module is in charge of identifying and flagging the exposures which are affected by a high stray light level or a high rate of cosmic ray impacts and when housekeeping temperatures are too high. The minimal angle values for a valid exposure are 120$^{\circ}$ for the Sun and 5$^{\circ}$ for the Moon. There is also a provided stray light estimate used to flag high stray light levels in the images.

	The ratio of bad exposures can be as high as 10 min per orbit on average for SAA and 40 min per orbit for the Earth occultation when the instrument line of sight is out of the ecliptic plane. Both types are not necessarily phased one over the other so they can overlap or happen at different time, lowering the duty cycle down to 50\% in the worst case. There could also be situations where there is only one or two valid exposures between two consecutive gaps that must be dealt with.

    Finally this module also verifies housekeeping temperature and checks for values that might lie outside predetermined bounds and could be responsible for bad measurements. The \DRP\ takes this information into account and flags each exposure accordingly.

	\subsection{Centroids}
    \label{ssec:centroid}
	An accurate reconstruction of the depointing is needed for some correction steps and for the photometry. The expected stability of the platform is about $\pm 2$ pixels on the long term. For each image, the on-board estimate of the depointing is the starting point of centroid determination.
	
	The centroid computation is applied to the image corrected from bad pixels and smearing. We use an iterative Gaussian apodization method (as in Deline et al. \textit{in press}). The algorithm starts by applying a Gaussian apodization on the target in order to reduce the influence of neighbouring stars, image corners and pixels entering and exiting the image with jitter. The Gaussian mask parameters are  $\sigma$=10 pix in relation with the PSF size, and the initial centering is the depointing provided by the on-board software. Then, the center of light is computed on the weighted image, resulting in a refined measurement. The mask is then re-centered accordingly and the process iterates until a convergence criterion is met. The convergence is usually obtained within 20 iterations. 

\begin{figure}
	   	\includegraphics[width=\linewidth]{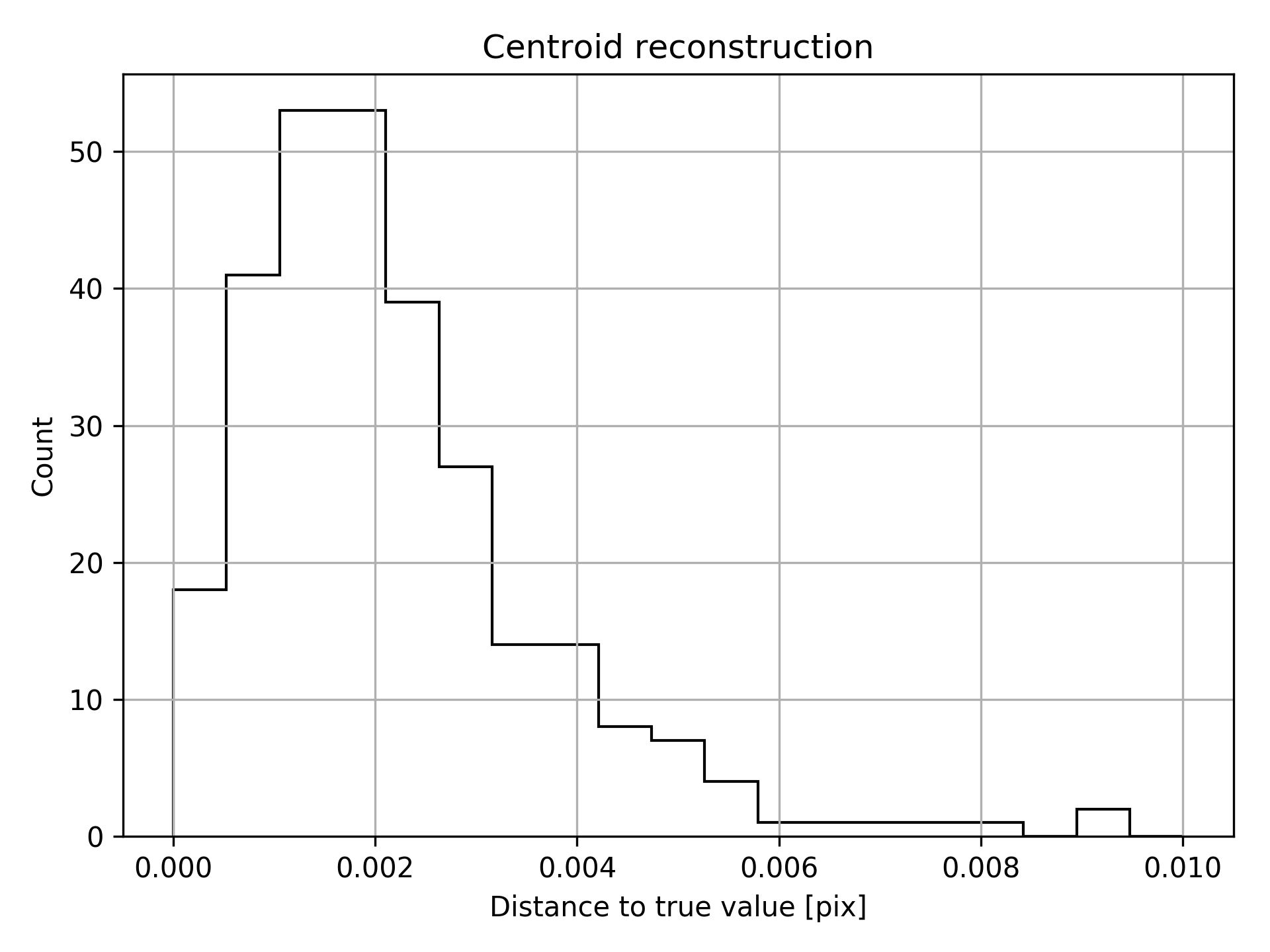}
		\caption{Distance between \DRP\ computed target's centroids to the \textit{true} values used in the simulations.}
		\label{fig:centroid}
	\end{figure}

	The centroid estimation error is as low as $2\times10^{-3}$~pix. Figure~\ref{fig:centroid} presents the distance from the estimate to the `true' depointing introduced in the simulations. The centroid of an image may differ by a constant between the methods of \DRP, on board software and reference PSF centering. To overcome this point, centroids around the center of the subarray are converted into depointings around (0, 0) by subtracting their own average. Thus, the result is consequently method independent.

\subsection{Pixel and sky coordinates relationship}
    \label{ssec:skycoord}

	In order to pass from pixel coordinates to physical sky coordinates back and forth, the \DRP\ uses the World Coordinate Systems (WCS) library \citep{Calabretta2002}. This library is commonly used in the astrophysics community as designed to easily store the sky coordinates in the data. Since the pipeline deals with several different images over the visit the WCS are stored in each individual image. The target has its coordinates RA and DEC defined. The centroid position in pixel coordinates and the rotation angle of each image define the reference point for WCS. The WCS rotation matrix is defined from the rotation angle coming from the raw data in order to be incorporated as WCS information. All necessary WCS keywords are stored in the metadata to be easily used with the WCS library to get the physical sky coordinates.

	\subsection{Report}
    \label{ssec:report}

    After each run of the \DRP\ an automatic report of the processed observation is generated in the form of a document provided to the  \Cheops\  end user. This report is a digest of plots and metrics that walks-through the gradual evolution of the signal across the successive \DRP\ steps. It is a fast way to identify possible noise sources in the final light curve, or any residual correlation that can exist between the target's flux and main observational parameters such as depointing, roll angle, etc. Usual metrics are point-to-point RMS, measurements of the scatter in some portions of the light curve, and a modified version of the Combined Differential Photometric Precision (CDPP) \citep{Jenkins2010b, Christiansen2012} to account for the gaps in the data. It is worth noticing that no filtering nor any detrending algorithm is applied in the metrics themselves in order to preserve the full signal information and, in this way, to accurately see its evolution. Examples of the report can be found in CHEOPS guest observer website at ESA\footnote{https://www.cosmos.esa.int/web/cheops-guest-observers-programme}.

\section{Performance}
\label{sec:performance}

Two datasets were prepared using CheopSim simulator (Futyan et al. \textit{subm.}) to illustrate the performance of the \DRP\ and compare it with the \Cheops\ science requirements in terms of photometric precision.

    The first simulation (case 1 hereafter) represents the observation of a transit of an Earth-size planet orbiting a $V$-mag=6 (G0V) star with a period of 50 days. 
    The second simulation (case 2 hereafter) corresponds to the observation of a faint star ($V$-mag=12) with a transit of a Neptune-size planet in a 13 days orbit. \Cheops\ photometric requirements for these science cases are 20 ppm in 6 hours of integration time for case 1; and 85 ppm in 3 hours of integration time for case 2. 
    
\begin{figure}
	   	\includegraphics[width=\linewidth]{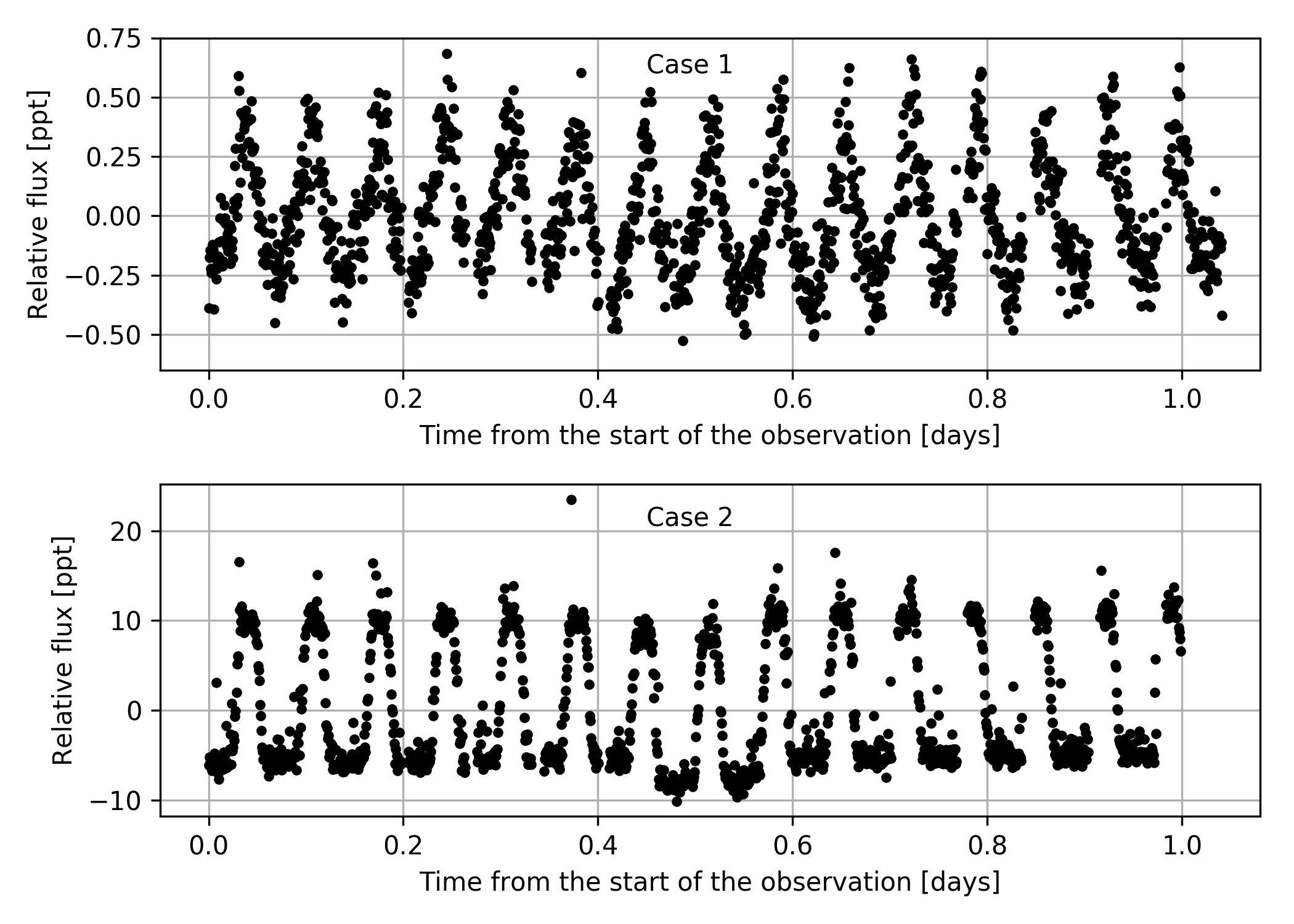}
		\caption{Light curves of raw data of case 1 (top) and case 2 (bottom).  The raw data has been only corrected for bias, dark and gain for units adaptation.}
		\label{fig:lcs_raw}
	\end{figure}

\begin{figure}
	   	\includegraphics[width=\linewidth]{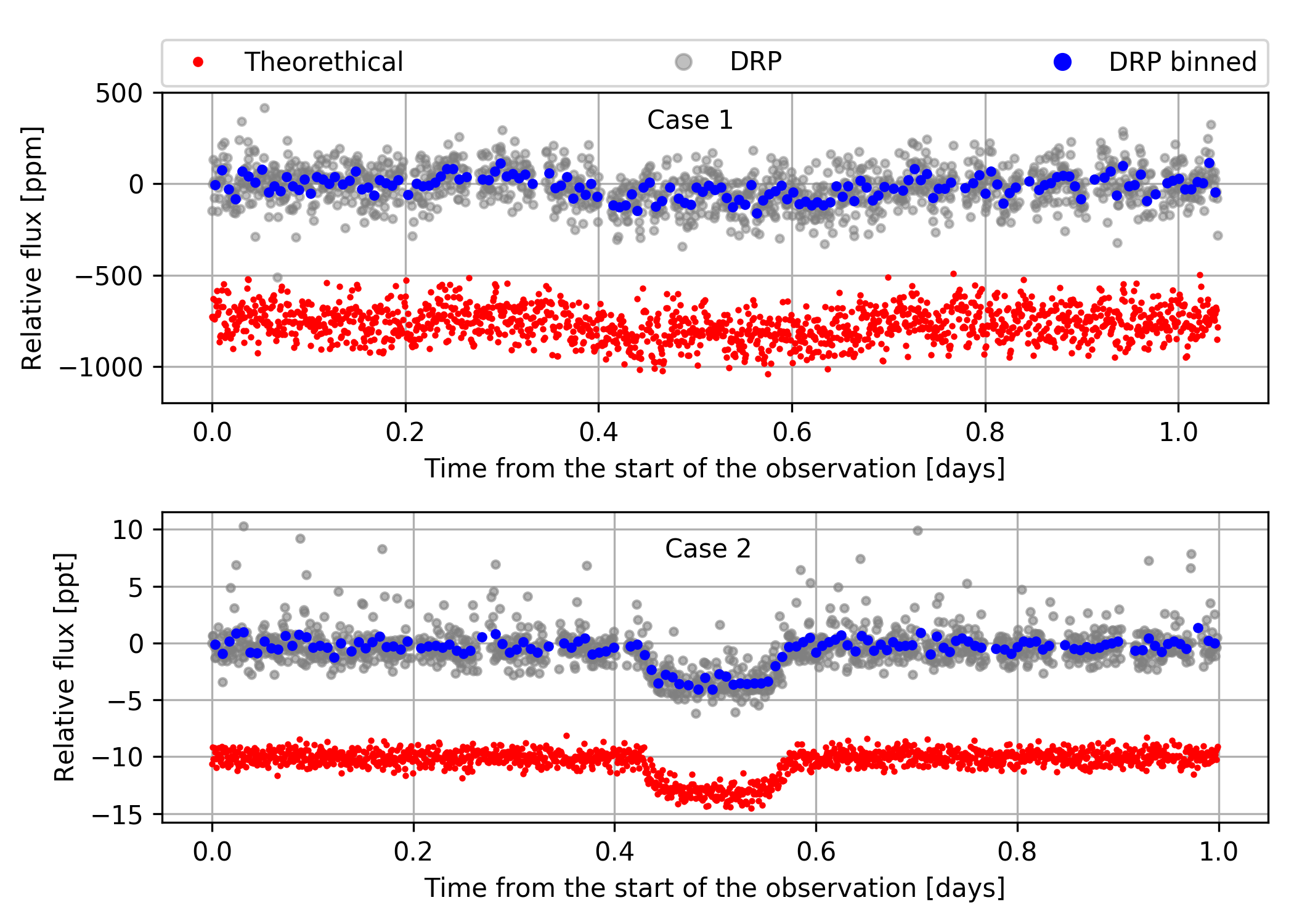}
		\caption{\DRP\ light curves with the default aperture (gray) in case 1 (top) and 2 (bottom). Blue points are the 10 min binned version. Red points are the unbinned theoretical light curve arbitrarily shifted for better visualization.}
		\label{fig:lcs_final}
	\end{figure}

    Both simulations were built using an intermediate contamination environment by setting the appropriate options of CheopSim to \textit{medium} (namely 1673 background stars in the field and $\sim2$ e$^{-}$ pix$^{-1}$ s$^{-1}$ of stray light) and taking into account the intrinsic stellar noise of its host star. Cosmic rays were randomly injected on both simulations together with 3 hot and 1 telegraphic pixels manually placed on the subarray window to avoid contaminating the target's PSF.  
    The light curves of both simulations are shown in Fig.~\ref{fig:lcs_raw}. The planet transits are barely visible in the simulator output light curves slightly processed by removing the  simulated bias, dark and gain to convert the units (hereafter \textit{raw} light curves). The cause is a strong correlation of flux with the position of the target on the CCD (case 1) and the background contamination in the aperture in case 2. The corresponding final \DRP\ default light curves and their 10 minutes binned version are shown in Fig.~\ref{fig:lcs_final}. In addition, the  theoretical light curves containing only photon and stellar noise, i.e. before the injection of any instrumental or environmental contamination, are also shown.     
    
    The modified CDPP at different time scales is used to asses the photometric precision of the light curves for each case.  It accounts for the gaps in the light curves and is based on the mean of the unbiased variance estimates from a rolling window of a specific time length. Then, the reported CDPP value corresponds to the square root of the mean of the variances normalized by the maximum number of points in the rolling windows. This metric can be interpreted as the noise one would obtain by rebinning the light curve at the selected time scale, time correlated or red noise included. As explained before, no detrending nor filtering is applied within the metrics.   
    
\begin{figure}
	   	\includegraphics[width=\linewidth]{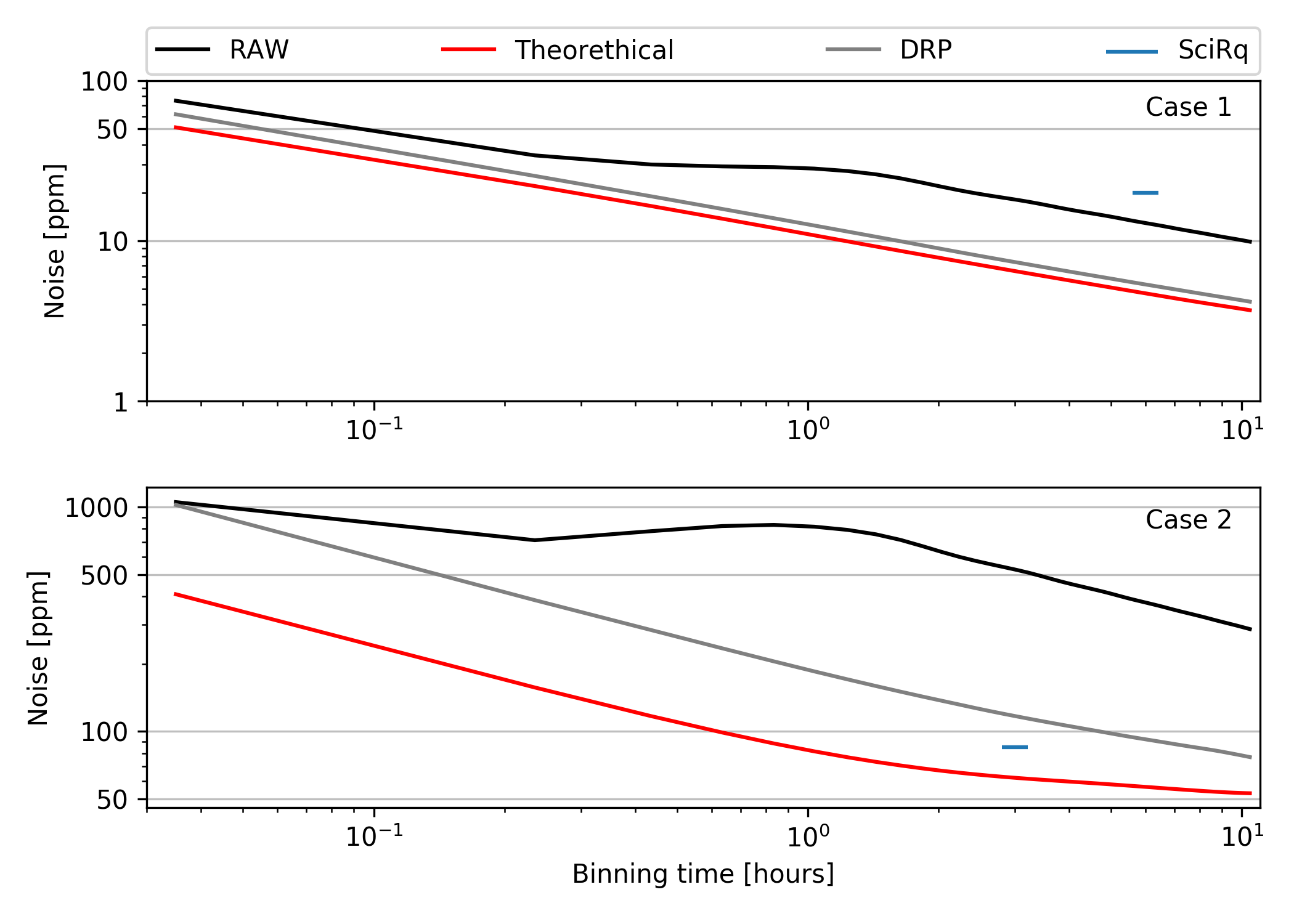}
		\caption{Noise estimations for the case 1 and 2 light curves. The plots are the modified CDPP of the raw (black), default \DRP\ (gray) and theoretical light curves. The photometric requirement for each case is represented by the blue dash at 6 and 3 hours for case 1 and 2, respectively (see text).}
		\label{fig:noise_curves}
	\end{figure}

    The obtained CDPP is shown at different time scales in Fig.~\ref{fig:noise_curves}. For case 1, the \DRP\ light curve reaches a precision below 6~ppm in 6~hours of integration time with a duty cycle of 90$\%$ in the 24 hours of the visit. 
    
    The noise estimation for case 2 is 117~ppm in 3~hours of integration for the same duty cycle. For its part, the optimal aperture (not shown in Fig.~\ref{fig:noise_curves}) delivers a dispersion of 83~ppm while the dispersion of the theoretical light curve is 62~ppm in the same conditions. The case 2 light curve is evidently affected by undetected cosmic rays  (Fig.~\ref{fig:lcs_final}). This effect is not surprising since the long exposure time (60 s) used in case~2 translates in a larger number of CR per image for an equivalent integrated flux (see for example Futyan et al. \textit{subm.}). Furthermore, in this observation no imagettes are available to help the cosmic rays detection. Confirming this effect, a control simulation with no cosmic rays injected gives dispersion of only 70~ppm (resp. 67~ppm  with the optimal aperture) comparable to the theoretical case of 62~ppm.    
    
    Regarding the detection of the hot and telegraphic pixels, the pipeline was able to recover both hot pixels out from a dark current 3 to 5 times larger than the usual one and the inserted telegraphic pixel was also correctly flagged. There were a few false detections, most of them close to the target, but they have no influence on the result since the \DRP\ is not correcting but communicating them for posteriori long term analysis.

    \begin{table}
        \centering
        \small\addtolength{\tabcolsep}{-3pt}
        \caption{CDPP estimations from the final light curves of the case 1 and 2 (see text for description).  The first column shows the integration time used for the metrics, the second column shows the photometric requirements for each case and the final columns shows the respective noise estimation of each analyzed light curve.}
        \label{tab:perf_cdpp}
        \begin{tabular}{cccccc}
            \hline
            Science & Int. Time & Req. & \multicolumn{3}{c}{Modified CDPP}\\
            Case & (hours) & (ppm) & Theor. & Default & Optimal \\
            \hline
            1     & 6    & 20   & 5   & 6 & 5 \\
            2     & 3    & 85   & 62  & 117 & 83\\
            2 w/o CR & 3  & 85    & 62 & 69      & 67 \\            
            \hline
        \end{tabular}
    \end{table}
\normalsize

	\section{Conclusions}
    \label{sec:conclusions}
    
    The \Cheops\ data reduction pipeline in its pre-launch version has been presented in this paper  with a detailed description of the core processing stages of the calibration, correction and photometry modules. The particularities of \Cheops\ data and their treatments by the pipeline have also been discussed. In addition, two representative examples of scientific cases for \Cheops\ have been used to evaluate the expected performance of the pipeline. For each case, the achieved photometric precision is given at different time scales. These examples show that the results of the \DRP\ are fully compliant with the scientific requirements of the mission. 
    %your point of view on final performance expectations, and the role of pipeline ?
    
    Even for challenging observations, such as for faint targets (e.g. case 2 in Sect.~\ref{sec:performance}) the light curve derived by the \DRP\ is not far from ideal results. It was shown that for a V-mag=12 target, the 3-h dispersion of the light curve derived with the optimal mask is very close to the noise level of the theoretical photometry: 83~ppm vs 62~ppm, respectively. In fact, this could even be improved by performing clipping of photometric outliers or flux binning, for example. These treatments are left to the users since best results are usually reached by a case by case detailed analysis which strongly depends on the science goals of the observations. As shown in Sect.~\ref{sec:performance}, the deviations from theoretical expected performance of \Cheops\, is not driven by instrumental effects but by the influence of external agents such as smear trails of background stars, cosmic rays and/or background contamination. The pipeline has proven that it is able to mitigate successfully these effects on the final photometry even though improvements, in particular, in cosmic ray detections are still under study and will be finally tested when the in-flight PSF is available.
    
    \DRP\ generates various output products the user will retrieve from the archive: four light curves calculated with different aperture sizes, each with its contamination curve and associated uncertainties. The user will also get the report automatically generated by the pipeline which intends to allow the user to follow what treatment has been done on the data, the quality of the processing and of the final result. In addition to these final products, the user will have the possibility to get additional by-products of the processing, such as, for example, bad pixel maps or the background light curve.
    
    %The examples of the performance presented here are only to illustrate that the pipeline is compliant with the requirements. But the pipeline performance has been intensively challenged on many more configurations, resulting in the internal validation by the mission and by ESA {\bf for me that's technical and should not go there}.
    
    After the launch, the pipeline will be tuned and adapted to real in-flight data: algorithms and modules will be improved all along the mission lifetime with our increasing knowledge and understanding of the instrument to allow the best characterization of the transiting planets \Cheops\ will observe. 
    
    The pipeline and its associated reference files are under versioning control and therefore the data can be easily re-processed if it is required by \Cheops\ project.

\begin{acknowledgements}
The \DRP\ team thanks our referee, F. Claus, for his very careful reading of the paper and his valuable comments and suggestions which help us to improve the quality of the manuscript.  We also thank D. Futyan for his valuable support on Cheopsim. We gratefully acknowledge the SOC team and the CHEOPS science team members who evaluated the pipeline results, for their constant support in the realization of the CHEOPS pipeline and their helpful suggestions which allowed significant improvements of key reduction steps. We thank K. Isaak for her careful reading and valuable feedback on the manuscript. We thank also A. Cameron and D. Queloz for providing comments that improved the paper. 
The team at LAM acknowledges CNES funding for the development of the \Cheops\ DRP, including grants 124378 for O.D. and 837319 for S.H., and the support of the Direction Technique of INSU with P.G. assignment. S.G.S also acknowledges support from FCT (FCT - Funda\c c\~ao para a Ci\^encia e a Tecnologia) through Investigador FCT contracts nr.IF/00028/2014/CP1215/CT0002. O.D. is also supported by FCT contract DL 57/2016/CP1364/CT0004. This work was supported by FCT/MCTES through national funds (PIDDAC) by the grant UID/FIS/04434/2019, by FCT funds (PTDC/FIS-AST/28953/2017) and by FEDER - Fundo Europeu de Desenvolvimento Regional through COMPETE2020 - Programa Operacional Competitividade e Internacionaliza\c c\~ao (POCI-01-0145-FEDER-028953). 

\end{acknowledgements}

\textit{Software:} The \DRP\ is developed in Python~3 (Python Software Foundation, https://www.python.org/), and makes use of Astropy \citep{astropy2013, astropy2018}, Matplotlib \citep{Hunter2007}, Numpy (https://www.numpy.org/), Scipy \citep{scipy} among other Python open source libraries. Jupyter notebooks \citep{Kluyver2016aa} were also used for the developing and testing of the code.

	%-------------------------------------------------------------------
	
\bibliographystyle{aa.bst}
\bibliography{DRP_CHEOPS_Hoyer}

\end{document}